\documentclass{aa}
\usepackage{natbib}
\bibpunct{(}{)}{;}{a}{}{,} 

\usepackage{graphicx}
\usepackage{txfonts}
\usepackage{upgreek}
\usepackage{comment}

\usepackage{tabularx} 
\usepackage{caption}
\usepackage{subfig}

\newcommand{\Kelvin}{\textrm{ K}}
\newcommand{\degrees}{^{\circ}}

\newcommand{\um}{\textrm{ } \upmu \textrm{m}}

\newcommand{\kms}{\textrm{ km } \textrm{s}^{-1}}

\newcommand{\plimit}{$0.12$}

\newcommand{\nsig}{$4.6$}
\newcommand{\epslimit}{$1.5 \times 10^{-5}$}
\newcommand{\epslimitone}{$0.50 \times 10^{-5}$}

\begin{document}

   \title{Searching for reflected light from $\tau$ Bootis b with high-resolution ground-based spectroscopy: Approaching the $10^{-5}$ contrast barrier}
   \titlerunning{Searching for reflected light from $\tau$ Bootis b}
	

   \author{H.J. Hoeijmakers \inst{1}, I.A.G. Snellen \inst{1}, S.E. van Terwisga \inst{1}}
	\authorrunning{Hoeijmakers et al.}
   \institute{1: Leiden Observatory, University of Leiden,
              Niels Bohrweg 2, 2333CA Leiden, The Netherlands\\
              \email{hoeijmakers@strw.leidenuniv.nl} \\}

   \date{Received 18 May 2017 / Accepted 6 September 2017}


  \abstract
   {It is challenging to measure the starlight reflected from exoplanets because of the extreme contrast with their host stars. For hot Jupiters, this contrast is in the range of $10^{-6}$ to $10^{-4}$, depending on their albedo, radius and orbital distance. Searches for reflected light have been performed since the first hot Jupiters were discovered, but with very limited success because hot Jupiters tend to have low albedo values due to the general absence of reflective cloud decks.}
   {The aim of this study is to search for reflected light from $\tau$ Boo b, a hot Jupiter with one of the brightest host stars. Since its discovery in 1997, it has been the subject of several reflected-light searches using high-dispersion spectroscopy. Here we aim to combine these data in to a single meta-analysis. 
}
  {We analysed more than 2,000 archival high-dispersion spectra obtained with the UVES, ESPaDOnS, NARVAL UES and HARPS-N spectrographs during various epochs between 1998 and 2013. Each spectrum was first cleaned of the stellar spectrum and subsequently cross-correlated with a PHOENIX model spectrum. These were then Doppler shifted to the planet rest-frame and co-added in time, weighted according to the expected signal-to-noise of the planet signal.}
  {We reach a 3$\sigma$ upper limit of the planet to star contrast of \epslimit. Assuming a planet radius of 1.15 $R_J$, this corresponds to an optical albedo of \plimit between 400-700 nm. This low albedo is in line with secondary eclipse and phase curve observations of other hot Jupiters using space-based observatories, as well as theoretical predictions of their reflective properties.}  
{}

   \keywords{
				Techniques: Spectroscopic	  
			    Planets and satellites: Atmospheres --
			    Planets and satellites: Detection --
			    Planets and satellites: Gaseous planets --
                Planets and satellites: $\tau$ Boo b
               }

   \maketitle
%

\section{Introduction}
Since the discovery of the first hot Jupiters, multiple attempts have been made to detect starlight reflected off their atmospheres using ground-based facilities \citep[e.g.][]{Charbonneau1999,Colliercameron1999,Colliercameron2000,Leigh2003,Rodler2010,Rodler2013}. However these studies have all resulted in non-detections and upper limits. To date, the only measurement of a ground-based measurement of reflected light of an exoplanet is claimed by \citet{Martins2015}, who report a detection of 51 Pegasi b.

A detection of direct reflected light can be used to constrain the reflectivity (albedo), as well as its spectral and directional dependence. This is an important observable because it is directly related to the global surface and/or atmospheric properties of the planet. This is exemplified by bodies in the solar system: lunar regolith appears dark at intermediate phase angles and preferably scatters light into either the forward or backward direction. The icy moons in the outer solar system have albedos close to 1.0, whereas rocky cometary nuclei may reflect only a few percent of the incoming light. Jupiter and Saturn have an orange hue due to organic hazes, while Uranus and Nepture are blue due to deep Rayleigh scattering combined with absorption by methane at red wavelengths \citep[see e.g.][]{Atreya1985,Wagener1986,Moses1995,Irwin2017} 

For the current generation of telescopes, the angular distances between  extrasolar planets and their host stars are generally too small to be spatially resolved. Moreover, the host stars are many orders of magnitude brighter than their planets, making these notoriously difficult to detect. Therefore in most cases, the presence of an orbiting planet needs to be inferred by measuring its indirect effect on the light of the host star. The transit and radial velocity methods have been the most successful in terms of yield \citep{Schneider2011}. These methods were first successfully applied by \citep{Charbonneau1999} and \citep{Mayor1995} respectively, and are mostly sensitive to gas giants in close-in orbits. Such hot Jupiters are relatively rare, but are overrepresented in the known exoplanet population because of this detection bias \citep[see e.g.][]{Howard2012,Fressin2013}. In principle, hot Jupiters are also the most favourable targets to search for reflected light: they intercept a relatively large fraction of the stellar flux due to their large radii and short orbital distances, potentially optimising the contrast between the planet and its star \citep{Charbonneau1999}. As follows from Appendix A, this contrast is a direct proxy for the albedo of the planet.

The albedo of a transiting hot Jupiter can readily be inferred from the depth of the secondary eclipse. Successful measurements are all attributed to space telescopes, such as Kepler, MOST, CoRoT, EPOXI and the Hubble space telescope \citep[see e.g.][and others]{Rowe2008,Christiansen2009,Alonso2009a,Alonso2009b,Santerne2011,Desert2011,Kipping2011,Coughlin2012,Esteves2013,Evans2013,Gandolfi2013,Morris2013,Demory2014,Shporer2015,Angerhausen2015,Gandolfi2015}. These have shown that hot Jupiters are generally dark, in line with theoretical models of the scattering properties of their atmospheres \citep[e.g.][]{Marley1999,Sudarsky2000,Burrows2008,Heng2013}. However, hot Jupiters with effective temperatures in the range of T$\sim$2000 - 3000 K can glow considerably at optical wavelengths, contaminating reflected starlight with intrinsic thermal emission. This has complicated the retrieval of the albedo from secondary eclipse measurements for the hottest planets \citep{Snellen2009,Snellen2010b,Cowan2011,Kipping2011,Demory2011,Esteves2013,Renyu2015}.

High-resolution spectroscopy offers an alternative way to discern the reflected planet light from the much brighter host star. Upon reflection by the planet, the starlight experiences a Doppler shift equal to the radial component of the planet's orbital velocity. As hot Jupiters generally have orbital velocities in excess of $\sim 100\kms$, these Doppler shifts are well resolved by modern Echelle spectrographs with spectral resolutions of $R=\frac{\lambda}{\Delta\lambda} > 50,000$. Using the known orbital parameters, spectra taken at any orbital phase can be shifted back to the rest frame of the planet and subsequently co-added in time. Over the past 20 years there have been a number of attempts to detect the reflected light of hot Jupiters in this way, but besides the detection of 51 Peg b claimed by \citet{Martins2015}, these have all resulted in upper limits on the planet-to-star contrast \citep[see e.g.][]{Charbonneau1999,Colliercameron2000,Colliercameron2002,Leigh2003,Leigh2003b,Rodler2008,Rodler2010}. The main results of these searches for reflected light using ground-based high-resolution spectroscopy are summarized in Table \ref{tab:previousresults}.

\begin{table*}
\begin{tabularx}{0.95\textwidth}{l|l|l|l|l|l|l|l}
\textbf{Planet} & \textbf{Reference} & \textbf{$\epsilon$} & \textbf{Confidence} & \textbf{Phase-function}& \textbf{$i$ ($K_p$)}& \textbf{$R_p$ ($R_J$)}  & \textbf{$p$}\\ 
\hline
$\tau$ Boo b	& \citet{Charbonneau1999}		& $<5\times10^{-5}$ & 			$  99\%$	  			& Lambertian & $\sim45\degrees$ & 1.2 	& <0.3\\
				& \citet{Colliercameron2000}	& $<3.5\times10^{-5}$ & 			$  99.9\%$ 				& Lambertian & $\sim40\degrees$ & 1.2	& <0.22 \\
				& \citet{Leigh2003}				& $<5.61\times10^{-5}$ & 		$  99.9\%$				& Venus-like & $36\degrees $ 	& 1.2	& <0.39 \\
				& \citet{Rodler2010}			& $<5.1\times10^{-5}$ & 			$  99.9\%$ 				& Venus-like & $46\degrees$		& 1.2	& <0.40 \\
\hline
 75289 b		& \citet{Leigh2003b}			& $<4.18\times10^{-5}$ & 		$  99.9\%$				& Venus-like & (127km/s) 		& 1.6	& <0.12\\
				& \citet{Rodler2008}			& $<6.7\times10^{-5}$ & 			$  99.9\%$				& Venus-like & (129km/s) 		& 1.2  	& <0.46\\
\hline
$\upsilon$ And b& \citet{Colliercameron2002}	& $<5.85\times10^{-5}$ & 		$  99.9\%$				& Venus-like & (135km/s) 		&	-	& -   \\
\hline
51 Peg b		& \citet{Martins2015}			& $6.0 \pm 0.4 \times10^{-5}$ & 		$3.7\sigma$				& Lambertian & (132km/s) 		&	$1.9 \pm 0.3$	& 0.5 	\\ 

\end{tabularx}
\caption{Results of previous high-resolution searches for reflected light of non-transiting hot Jupiters. The last four columns indicate the assumed phase function, the best-fit orbital inclination $i$ (or equivalent: the best-fit semi-amplitude $K_p$ of the radially projected orbital velocity) and the inferred limiting combination of planet radius and grey albedo. The works that use Venus-like phase functions use an empirical model formulated by \citet{Hilton1992}. The only work in this list that claims an actual measurement of $\epsilon$ is the work by \citet{Martins2015}. All other studies report upper limits. The upper limit by \citet{Colliercameron2000} was adjusted by \citet{Leigh2003} and is therefore superseded by it. The limit quoted from \citet{Charbonneau1999} is inferred from the inclination-dependent contrast curve presented in their work.}  
\label{tab:previousresults}
\end{table*}

$\tau$ Bo\"otis b was discovered in 1997 along with two other hot Jupiters \citep{Butler1997}. It orbits a bright $(V_{\textrm{mag}}=4.5)$ F8 main-sequence star located 15.6pc from Earth, but was found not to transit \citep{Baliunas1997}. This star is one of the most metal-rich exoplanet hosts, and is orbited by a resolved M-dwarf companion at a distance of 240 AU \citep{Hale1994,Patience2002}. The properties of the system are listed in Table \ref{tab:systemparameters}. 

\begin{table}
\begin{tabularx}{0.5\textwidth}{l|l|l}
\textbf{Parameter} & \textbf{Sym.} & \textbf{Value} \\ 
\hline

Visible magnitude$^a$					& $V$ 					& $4.50$ \\ 
Distance $\textrm{(pc)}^b$ 				& $d$ 					& $15.60 \pm 0.17$ \\
Effective temperature $(\textrm{K})^c$ 	& $T_{\textrm{eff}}$ 	& $6399 \pm 45 $ \\
Luminosity $(L_{\odot})^c$  			& $L_*$ 				& $3.06 \pm 0.16$ \\
Mass $(M_{\odot})^c$ 		   			& $M_*$ 				& $1.38 \pm 0.05$ \\
Radius $(R_{\odot})^c$ 					& $R_*$ 				& $1.42 \pm 0.08$ \\
Surface gravity $\textrm{(cgs)}^d$ 		& $\log g$ 				& $4.27^{+0.04}_{-0.02}$ \\
Systemic velocity $\textrm{(}\kms\textrm{)}^e$ & $\gamma$ 		& $-16.54 \pm 0.34$ \\
Metallicity $(\textrm{dex})^{c}$		& $\left[\textrm{F}/ \textrm{H} \right]$  & $0.26 \pm 0.03$ \\
Age $\textrm{(Gyr)}^c$					& 						& $0.9 \pm 0.5$ \\ 
Rotation velocity $\textrm{(}\kms\textrm{)}^c$ & $v\sin (i)$      & $14.27 \pm 0.06$ \\

\hline

Orbital period $\textrm{(days)}^c$             	& $P$               & $3.3124568$ \\
												&					& $\pm 0.0000069$ \\
Semi-major axis $\textrm{(AU)}^c$     		   	& $a$               & $0.049 \pm 0.003$ \\
Orbital inclination $\textrm{(deg)}^f$       		   	& $i$               & $44.5 \pm 1.5$ \\
$\textrm{Eccentricity}^c$    	  				& $e$				& $0.011 \pm 0.006$ \\
Mass $(M_{\textrm{J}})^c$                       & $M_p$             & $6.13 \pm 0.17$ \\
Phase zero-point $\textrm{(HJD)}^f$			   	& $T_0$             & $2,455,652.108 \pm 0.004$ \\
\end{tabularx}
\caption[lalaaa]{Properties of the star $\tau$ Boo (upper part) and $\tau$ Boo b (lower part). \\ %

$a:$ Adopted from \citet{Valenti2005}.\\
$b:$ Adopted from \citet{Belle2009}.\\
$c:$ Adopted from \citet{Borsa2015}. \\
$d:$ Adopted from \citet{Takeda2007}. \\
$e:$ Adopted from \citet{Nidever2002}. \\
$f:$ Adopted from \citet{Brogi2012}. \\
}  
\label{tab:systemparameters}
\end{table}
   
$\tau$ Boo was observed by the MOST spacecraft for 37.5 days in 2004 and 2005 to monitor its variability. Searching this data for evidence of planet-to-star interaction, \citet{Walker2008} found a variable region on the stellar surface that is synchronized with the orbital period of the planet, and is likely magnetically induced. From spectro-polarimetric observations, \citet{Donati2008b} were able to map the magnetic field of the star and found evidence for differential rotation, with the rotation period varying from 3.0 to 3.9 days from equator to pole, consistent with the orbital period of the planet. This synchronization of star and planet has important consequences for the reflection spectrum. As seen from Earth, the stellar spectrum is rotationally broadened with $v\sin (i) = 15\kms$. However, since the planet co-rotates with the stellar surface, the star does not rotate from the perspective of the planet. Therefore, the reflected light spectrum is not expected to be strongly rotationally broadened, resulting in the absorption lines being significantly more narrow than those in the stellar spectrum as observed from Earth. The reflected stellar absorption lines are only broadened by the axial rotation $(v \sin (i) = 1.24\kms)$ of the planet itself \citep{Rodler2010,Brogi2013}. Techniques that aim to detect the absorption lines of the planet are therefore particularly sensitive for this system, owing to the fact that the planet and the star are tidally locked.

Because $\tau$ Boo is such a bright star, efforts to detect the reflection spectrum of the planetary companion quickly commenced after its discovery in 1997. The first search for reflected light was performed by \citet{Charbonneau1999}, who observed the system for three consecutive nights with the HIRES spectrograph at the 10 m Keck Observatory, between 465.8 nm and 498.7 nm. By fitting a scaled, Doppler-shifted version of a stellar template to each individual spectrum, they constrained $\epsilon$ to be less than $5 \times 10^{-5}$ to $8 \times 10^{-5}$, depending on the orbital inclination which was unknown at the time. Assuming a radius of $1.2 R_J$, the albedo was constrained to $p < 0.3$. \citet{Colliercameron1999} analysed 48 hours of high-resolution spectra obtained by the now decommissioned Utrecht Echelle Spectrograph (UES) at the 4.2 m William Herschel Telescope on La Palma, Spain. Following a similar procedure as \citet{Charbonneau1999}, they detected a signal equivalent to $\epsilon = 1.9 \pm 0.4 \times 10^{-4}$ between 456 nm and 524 nm. These authors explained the discrepancy between their results and those of \citet{Charbonneau1999} by the assumption of a different phase-function, differences in their template fitting procedure and a different way of treating systematics. However, after obtaining more data with the same instrument in early 2000, they were unable to reproduce this signal, this time constraining $\epsilon$ to $3.5 \times 10^{-5}$ \citep{Colliercameron2000}. Again assuming a planetary radius of $R_p=1.2 R_J$, this resulted in an albedo of $p < 0.22$. In 2003, the same group combined and re-analysed all data they obtained from 1998 to 2000 with the UES, and adjusted their $3\sigma$ upper limit to $\epsilon < 5.61 \times 10^{-5}$, assuming an orbital inclination of $i=39\degrees$ \citep{Leigh2003}.

In 2007, \citet{Rodler2010} obtained two nights of high-resolution spectra with the UVES spectrograph at the UT2 of the VLT. Following the fitting method of \citet{Charbonneau1999}, their analysis constrained the planet-to-star contrast ratio to $5.1 \times 10^{-5}$ to $5.7 \times 10^{-5}$ between 425 to 632 nm, depending on the assumed wavelength dependence of the albedo, and assuming an orbital inclination of $i=60\degrees$ and planet radius of $R_p=1.2R_J$.

In 2011, \citet{Brogi2012} and \citet{Rodler2012} independently observed $\tau$ Boo with the CRyogenic InfraRed Echelle Spectrograph (CRIRES) mounted on UT1 at the VLT, both targeting the CO absorption band at $2.3\um$ in the planet's intrinsic thermal spectrum. By cross-correlating their spectra with a CO model template, both groups were able to significantly retrieve the planet's CO absorption at $6\sigma$ and $3.4\sigma$ confidence respectively. The Doppler shift of this CO spectrum revealed the radially projected velocity semi-amplitude $K_p$ of the planet. Together with the known orbital velocity $v_{\textrm{orb}}$ and under the assumption of a circular orbit, this directly yields the orbital inclination $i$:

\begin{equation}\label{Eq:Kp}
K_p = v_{\textrm{orb}} \sin (i) = \frac{2 \pi a}{P} \sin (i).
\end{equation}

These observations thus constrained the orbital inclination of $\tau$ Boo b to $i = 44.5 \pm 1.5\degrees$ and  $i = 47^{\circ+7}_{-6}$ respectively. For the first time, a non-transiting hot Jupiter was detected directly, and the degeneracy between $M_p$ and $i$ that is inherent to the radial velocity method could be broken. In the case of a small spin-orbit misalignment angle, the stellar rotation velocity of $v\sin(i)=15\kms$ is consistent with an orbital period of 3.3 days, providing an independent confirmation of synchronization of the stellar rotation with the orbital period of the planet \citep{Brogi2012}. \citet{Rodler2013b} proceeded to re-analyse their UVES data with the known orbital inclination and an updated set of orbital parameters. Using the same fitting procedure as used in \citet{Rodler2012}, they were not able to significantly detect the reflection spectrum of the planet.

In this paper, we perform a meta search for reflected light from $\tau$ Boo b by combining data from the UVES, ESPaDOnS, NARVAL UES and HARPS-N spectrographs during various epochs between 1998 and 2013. Section 2 describes the observations and data analysis. In Section 3 we present our results and possible caveats, the implications of which are discussed in Section 4.

\section{Observations}
\begin{figure*}
   \centering
   \includegraphics[width=0.9\linewidth]{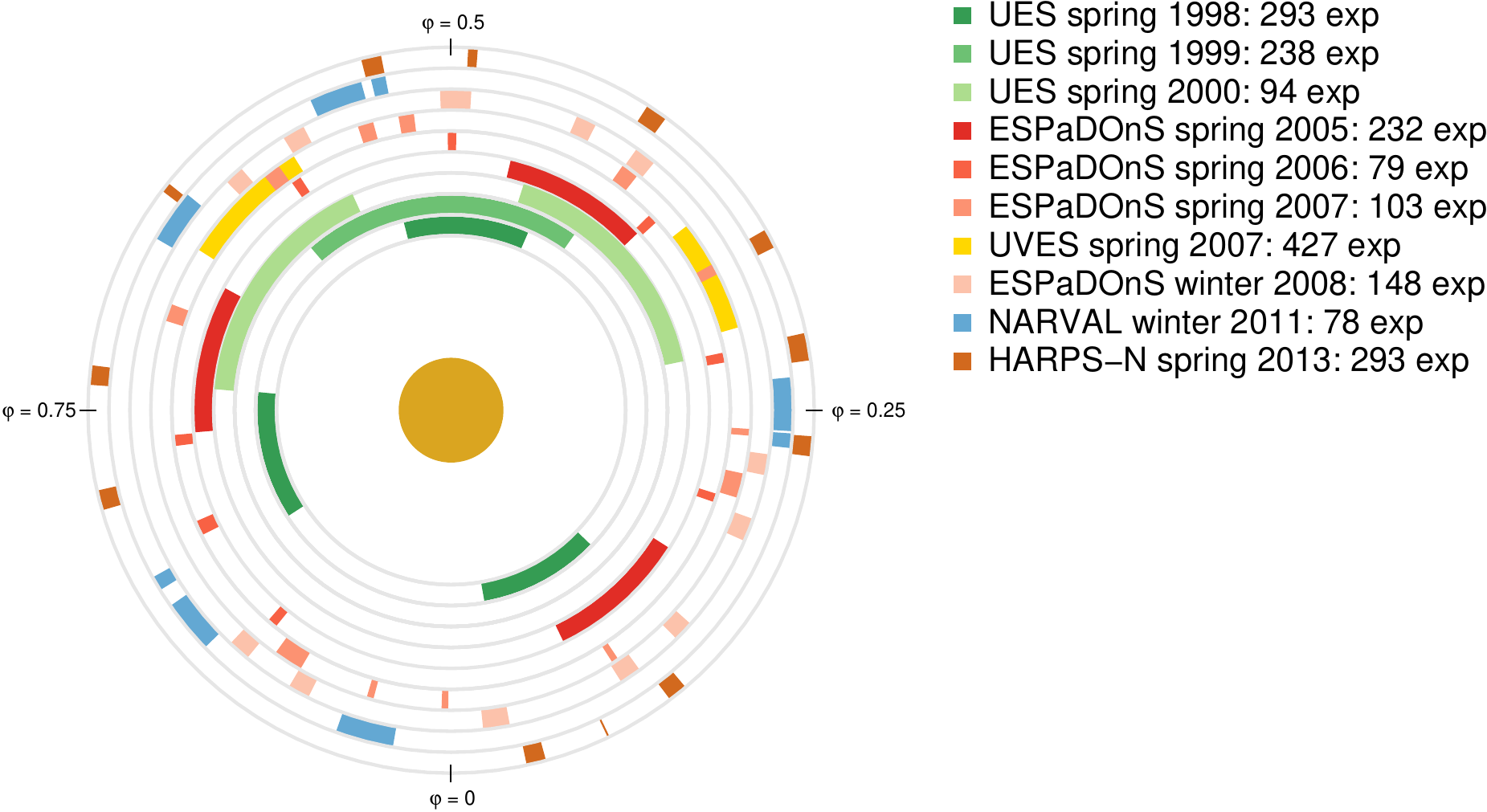}
   \caption{Diagram of the distribution of orbital phases covered by the observations used in this work. The data are grouped into rings, with the earliest data in the inner ring, moving outward chronologically. This diagram shows the varying nature of the observing strategies used for the different programmes. The UES obervations of 1998-2000, the ESPaDOnS observations of 2005 and the UVES observations of 2007 targeted the system consecutively for multiple hours, while the other ESPaDOnS, NARVAL and HARPS-N spectra were used to monitor the system periodically over longer periods of time. The UES and UVES observations were mostly obtained at large phase angles because these observations specifically targeted the planet's day side. 
}%
\label{fig:phase_coverage}
\end{figure*}

$\tau$ Boo has been observed as part of several observing campaigns, and previous studies have shown that reflected light from the planet 
is too faint to be detected in a single night of data. In this analysis we therefore combine the archival high-resolution, high-signal-to-noise optical spectra obtained with the UVES spectrograph at the 8-m VLT/UT2 \citep{Dekker2000}, the ESPaDOnS spectropolarimeter at the 3.6-m CFHT \citep{Donati2003}, the NARVAL spectropolarimeter at the 2-m TBL \citep{Auriere2003}, the UES spectrograph at the 4-m WHT \citep{Walker1986} and the HARPS-N spectrograph at the 3.6-m TNG \citep{Cosentino2012}. These contain the vast majority of all high-resolution observations of this system, except those that form the basis of \citet{Charbonneau1999} because these HIRES data were not preserved digitally and are likely lost (David Charbonneau, private communication). Table \ref{tab:datasets} presents an overview of all 25 datasets that are analysed in this work, with Fig. \ref{fig:phase_coverage} showing the orbital phase coverage of all of the data.

\subsection{UES data of $\tau$ Boo}\label{sec:UES}
Soon after the discovery of $\tau$ Boo b, \citet{Colliercameron2000} initiated an observing campaign to search for the reflected light of $\tau$ Boo b using the Utrecht Echelle Spectrograph mounted on the William Herschel Telescope at the Roche de Los Muchachos observatory on La Palma. They observed the $\tau$ Boo system for 17 complete and partial nights during three observing seasons in the years 1998, 1999 and 2000, obtaining 893 spectra in total\footnote{These observations are described in detail by \citet{Leigh2003}.}. The raw data are available in the online data archive of the Isaac Netwon Group of telescopes \footnote{http://casu.ast.cam.ac.uk/casuadc/ingarch/query} and were reduced in a standard way using IRAF version 2.16.1. In the analysis we treat each night of observations as an independent dataset, and use 63 to 67 spectral orders\footnote{Depending on the signal-to-noise of the bluest and reddest orders.} ranging from 407 nm to 649 nm at a spectral resolution of $\textrm{R}=53,000$. During the data reduction, we discarded 60 of the 893 spectra that were visibly affected by poor observing conditions, including the entire nights of April 13, 1998 and June 4, 1999. In total, the remaining data consist of 68.54 hours of observations.

\subsection{ESPaDOnS \& NARVAL data of $\tau$ Boo}\label{sec:Espadons}
\begin{figure*}
   \centering
   \includegraphics[width=\linewidth]{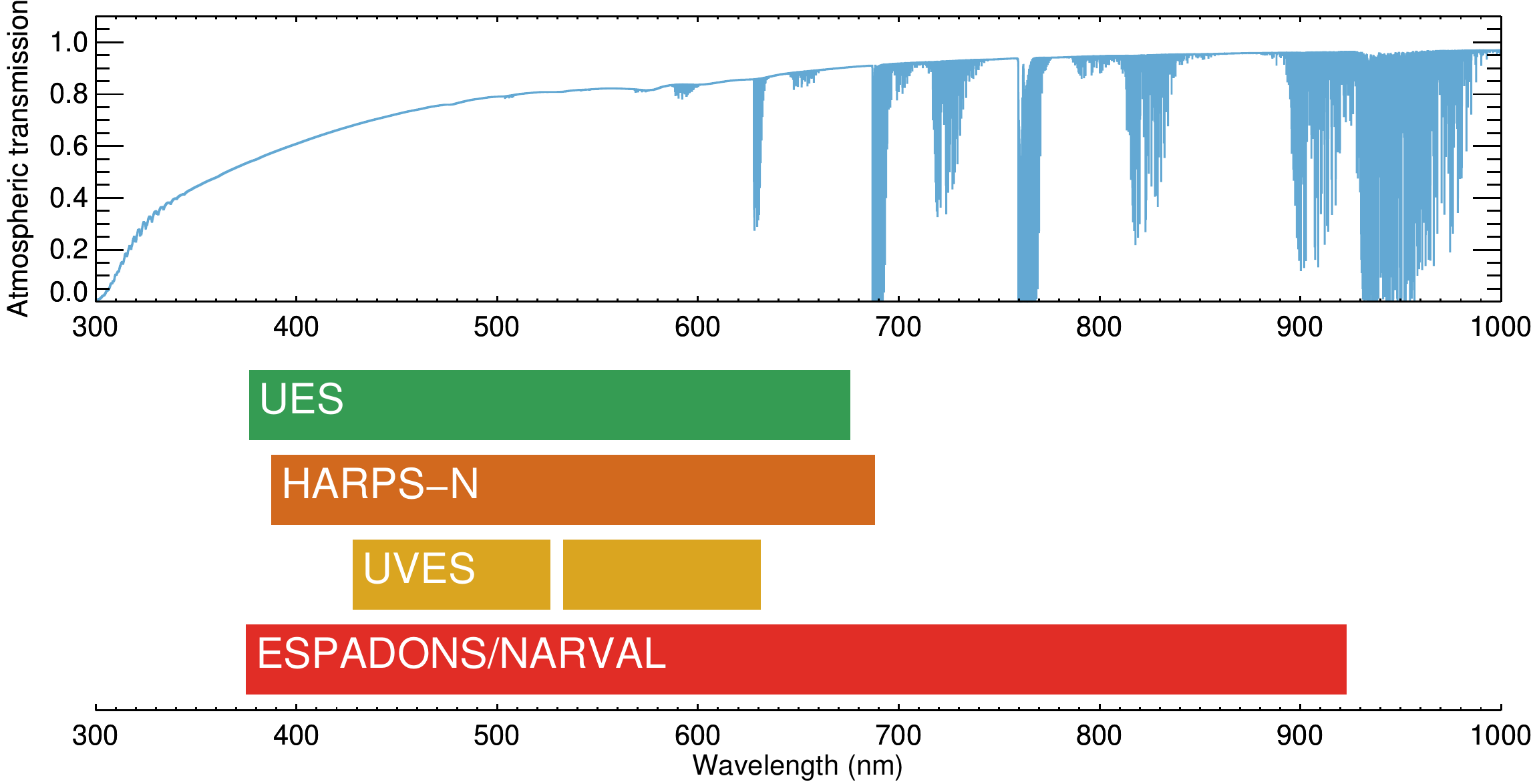}
   \caption{Wavelength coverage of the data. The top panel indicates the transmission function of the Earth's atmosphere, and the bottom panel shows the wavelength coverage of the data from the four instruments used in this work. Telluric contamination adversely affects our analysis, but regions that contain telluric lines are down-weighted by our analysis procedure according to the effect of the telluric lines on the cross-correlation function (see Section \ref{Postproc}).}%
\label{fig:Orders}
\end{figure*}

The NARVAL and ESPaDOnS spectro-polarimeters are nearly identical fiber-fed bench-mounted echelle spectrographs, located at the 3.6-m Canadian French Hawaiian Telescope and the 2-m Telescope Bernard Lyot on Pic du Midi in the French Pyrenees respectively. 
Reduced observations of $\tau$ Boo obtained by the ESPaDOnS and NARVAL instruments are publicly available via the Polar Base online database \citep{Petit2014}\footnote{http://polarbase.irap.omp.eu/}. The data were originally taken to study time variations in the magnetic field of the star \citep{Donati2008b,Fares2009}. We downloaded all 751 polarization spectra that were taken during multiple programmes between March 2005 and January 2011. Querying the Polar Base database for $\tau$ Boo returns a total of 971 spectra. 751 of these are individual polarization channels, while 220 are intensity channels that are obtained by combining multiple polarization channels and thus contain no independent information. All spectra have a wavelength coverage from 369.45 nm to 1048.5 nm at a spectral resolution of $\textrm{R}=65,000$, with exposure times between 150 and 600 seconds, resulting in a continuum signal-to-noise between 100 and 1000 per pixel. Although each exposure is formatted as a single 1D spectrum, the 40 individual echelle orders are not stitched together and were retrieved from the downloaded reduced data files. The wavelength coverage of the data is shown in Figure \ref{fig:Orders}.

Of the 751 spectra in total, 229 were taken on three consecutive nights in March 2005, while the other 522 spectra were obtained at various time intervals between June 2006 and January 2011, containing between 1 and 25 exposures at a given night. The three consecutive nights obtained in 2005 were treated as three individual datasets. The automatic pipelines used to reduce the data has caused the presence of negative values in some orders at the edges of the waveband (orders 1-3 and 37-39). Any spectral orders with such artefacts were discarded from the analysis. We chose to split the groups of 
ESPaDOnS + NARVAL spectra into four separate sets which were analysed independently of each other (each containing observations taken within a period of approximately one month), because our analysis relies on the time-stability of the stellar spectra (see Section \ref{Postproc}). 
We chose to disregard 131 exposures that were obtained sparsely over longer periods of time. In total, the data used in our analysis consist of 49.85 hours of observations.

\subsection{UVES data of $\tau$ Boo}\label{sec:UVES}

$\tau$ Boo was observed with UVES \citep{Dekker2000} on June 16 and 17, 2007 by \citet{Rodler2010}\footnote{ESO programme 079.C-0413(A). Data obtained from the ESO Science Archive Facility.}. These consist of 422 1D spectra covering a wavelength range between 427.88 nm and 630.72 nm at a resolution of $R=110,000$. The individual echelle orders were stitched together by the data reduction pipeline, hence each downloaded spectrum covers the full bandwidth of the instrument set-up. We chose to slice the spectra into 21 band-passes (hereafter referred to as 'orders', even though these slices technically do not exactly match the echelle orders of the UVES spectrograph) that we analyse independently from each other.

The two nights cover orbital phases between $0.29 < \varphi <0.35$ and $0.59 < \varphi < 0.66$ respectively (by convention $\varphi$ varies between 0 and 1, with inferior conjunction as zero point). This timing was chosen by the observers to maximize the fraction of the day-side of the planet in view, while ensuring that the planet has an appreciable radial velocity, allowing the reflected spectrum to be discerned from the spectrum of the star through its relative Doppler-shift.

\subsection{HARPS-N data of $\tau$ Boo}\label{sec:HARPS-N}
The $\tau$ Boo system is being monitored using the HARPS-N spectrograph at the 3.6-m Telescopio Nationale Galileo at Roche de los Muchachos Observatory on La Palma as part of the GAPS programme to characterize known exoplanet host stars and search for additional planetary companions using the standard radial velocity technique, astroseismology and the Rossiter-McLaughlin effect \citep[see e.g.][]{Covino2013,Desidera2013,Sozzetti2013}. To date, the TNG data archive includes 531 exposures of the $\tau$ Boo system. The first 285 of these were publicly available and were obtained in the second half of April and the first half of May of 2013. These were initially analysed by \citet{Borsa2015} to investigate the host star, the orbit of the outer stellar companion $\tau$ Boo B, and to update the orbital ephemeris of the planet.

The reduced 1D spectra were downloaded from the TNG public data archive\footnote{http://ia2.oats.inaf.it/archives/tng} and cover wavelengths from 387 nm to 691 nm at a spectral resolution of $R\sim110,000$. As with the UVES data, we sliced the spectra into 21 bands that were analysed separately. Because of the highly stable nature of the HARPS-N spectrograph, we chose to group the entire time-series together and treat it as a single set. In total, these data consist of 5.14 hours of observations.

\begin{table}
\begin{tabularx}{0.5\textwidth}{l|l|l|l|l|l}
\textbf{Set} & \textbf{Instrument} & \textbf{$N_\textrm{exp}$} & \textbf{$t_\textrm{exp}$ (hr)}&\textbf{Epoch} & \textbf{$\varphi$ range}\\ 
\hline

1	& UES		& 99  	&3.86& 09-04-1998    & 0.42-0.53\\ 
2	& UES		& 113 	&3.76& 10-04-1998	& 0.72-0.83\\
3	& UES		& 81  	&4.20& 11-04-1998 	& 0.03-0.13\\ 
4	& UES		& 45 	&4.22& 02-04-1999 	& 0.50-0.60\\
5	& UES		& 39 	&3.96& 25-04-1999 	& 0.44-0.54\\ 
6	& UES		& 61 	&4.20& 05-05-1999 	& 0.46-0.55\\ 
7	& UES		& 48 	&5.24& 25-05-1999 	& 0.48-0.58\\
8	& UES		& 44 	&3.56& 28-05-1999 	& 0.39-0.47\\ 
9	& UES		& 47 	&5.44& 14-03-2000 	& 0.27-0.37\\ 
10	& UES		& 44 	&6.10& 15-03-2000 	& 0.57-0.67\\ 
11	& UES		& 29 	&3.22& 24-03-2000 	& 0.29-0.39\\
12	& UES		& 44 	&5.35& 23-04-2000 	& 0.32-0.42\\ 
13	& UES		& 56 	&5.19& 24-04-2000 	& 0.62-0.73\\
14	& UES		& 41	&4.83& 13-05-2000 	& 0.36-0.44\\ 
15	& UES		& 42 	&5.39& 17-05-2000 	& 0.56-0.65\\
16	& ESPaDOnS	& 76  	&6.11& 23-03-2005 	& 0.07-0.16\\ 
17	& ESPaDOnS	& 76  	&6.33& 24-03-2005 	& 0.37-0.46\\
18	& ESPaDOnS	& 77  	&6.22& 25-03-2005 	& 0.67-0.76\\ 
19	& ESPaDOnS	& 75  	&3.01& 06-2006    	& Variable\\
20	& UVES		& 105 	&3.88& 16-06-2007 	& 0.29-0.35\\ 
21	& UVES		& 317 	&3.77& 17-06-2007 	& 0.59-0.66\\
22	& ESPaDOnS	& 103 	&7.02& 07-2007    	& Variable\\ 
23	& ESPaDOnS	& 142 	&9.33& 01-2008    	& Variable\\
24	& NARVAL	& 71  	&11.83& 01-2011    	& Variable \\
25  & HARPS-N	& 285 	&5.14& 05-2013		& Variable\\

\end{tabularx}
\caption[lalaaa]{The 25 datasets analysed in this work in chronological order, showing the instrument used, the number of exposures, amount of time spent on target, the observing epoch, and the range in orbital phase covered.}  
\label{tab:datasets}
\end{table}

\section{Data analysis}
\subsection{Post processing}\label{Postproc}
The reflected planet spectrum is expected to be present in the data at a level of $\epsilon \sim 10^{-5}$ times the stellar spectrum, assuming a planet radius and albedo in the order of $1R_J$ and $10\%$ respectively (see Eq. \ref{Eq:Geometric_albedo}). The high-resolution spectra typically have a peak signal-to-noise of $\sim 500 - 1000$ per pixel. The sensitivity of the observations is subsequently enhanced by three orders of magnitude by combining the signal from $\sim$10$^3$ individual absorption lines and the $\sim$2100 spectra. This procedure is similar to that used in previous works that use cross-correlation at high spectral resolution \citep[e.g.][]{Charbonneau1999,Colliercameron2000,Leigh2003,Snellen2010,Brogi2012,Rodler2012,Hoeijmakers2015}. First, the stellar absorption lines need to be removed from the spectra to reveal the faint Doppler shifted copy originating from reflection by the planet. For each data set, we obtain the stellar spectrum by time-averaging the spectra, which is subsequently subtracted from the data. This procedure removes all time-constant spectral features, but not the planet's reflected spectrum because its radial velocity changes with up to $8.4\kms$ per hour ($\sim$3$\times$ the full width at half maximum (FWHM) of the line-spread function at $R=100,000$). Residual time-dependent features arise from changes in the stellar line-shapes caused by variations in spectral resolution due to weather and seeing, telluric absorption, and stellar chromospheric activity. Therefore a significant part of our analysis focuses on their removal. The residual spectra are subsequently cross-correlated with a model template spectrum of the host star Doppler-shifted to the rest frame of the planet, which are then co-added in time. 

\begin{figure*}
   \centering

   \includegraphics[width=1.0\linewidth]{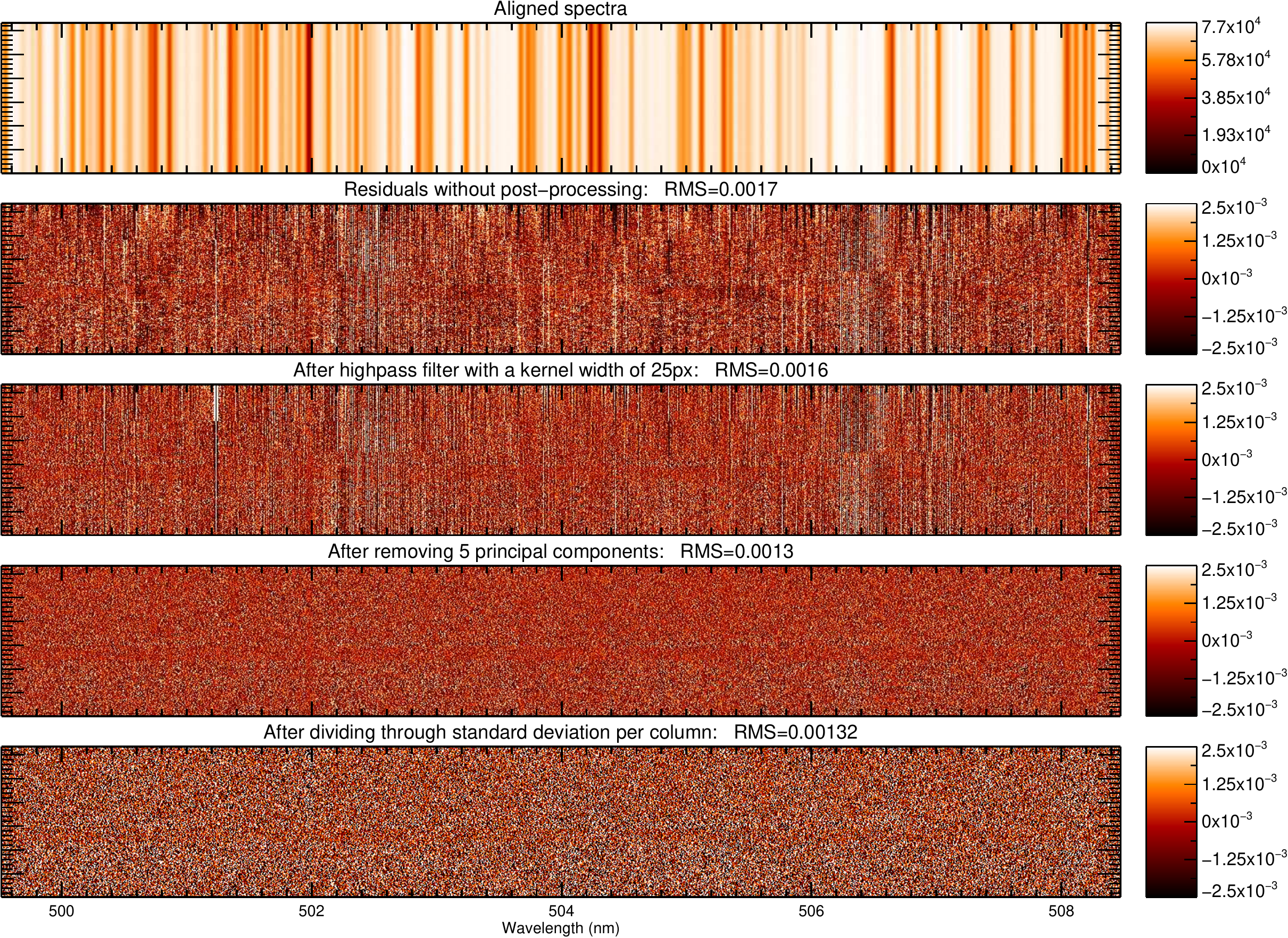}
   \caption{Step-wise analysis of order eight of the UVES observations obtained during the night of June 17, 2007. \textbf{First panel:} The aligned spectra (after step 1). \textbf{Second panel:} The residuals after subtracting the mean value from  each column. \textbf{Third panel:} The residuals after applying a high-pass box-filter with a width of 40 pixels (roughly $40\kms$). \textbf{Fourth panel:} After subsequent removal of four principal components (step 5). \textbf{Fifth panel:} After dividing each column from the fourth panel by the standard deviation in each column (step 6). This weighs down columns that intrinsically have a lower signal-to-noise. The strong vertical structures in the third panel are telluric water lines. These are effectively removed by the principal component analysis.}%
\label{fig:postproc}
\end{figure*}

Below is a description of the sequence of processing steps in detail, which are executed for each order of each dataset independently. An example of the step-wise analysis of a single spectral order is shown in Figure \ref{fig:postproc}.

	\begin{enumerate}
		
\item[1] \textbf{Alignment of spectra:} Instabilities in tracking, weather and a changing radial velocity of the observatory with respect to $\tau$ Boo cause the spectra to drift in wavelength over the course of an observing run\footnote{For UVES in particular, instrumental instability may cause velocity shifts in the order of $1\kms$. See \citet{Czesla2015} for an extreme example in the case of transit observations of HD 189733 b.}. Such velocity variations must be removed before the time-averaged stellar spectrum can be obtained. For this purpose, we identify all stellar absorption lines stronger than $6.0\%$ in the individual spectra, and locate their centroids by fitting a Gaussian line profile to the core of each line. Telluric lines are identified and rejected using a model telluric absorption spectrum obtained from ESO's SkyCalc Sky model calculator \citep{Noll2012,Jones2013}. We used these fitted line positions to align all spectra to a common reference frame. Fig. \ref{fig:Alignment} shows the average shift needed to align each exposure of one UVES, ESPaDOnS and UES night.

 \begin{figure*}
   \centering
   \includegraphics[width=1.0\linewidth]{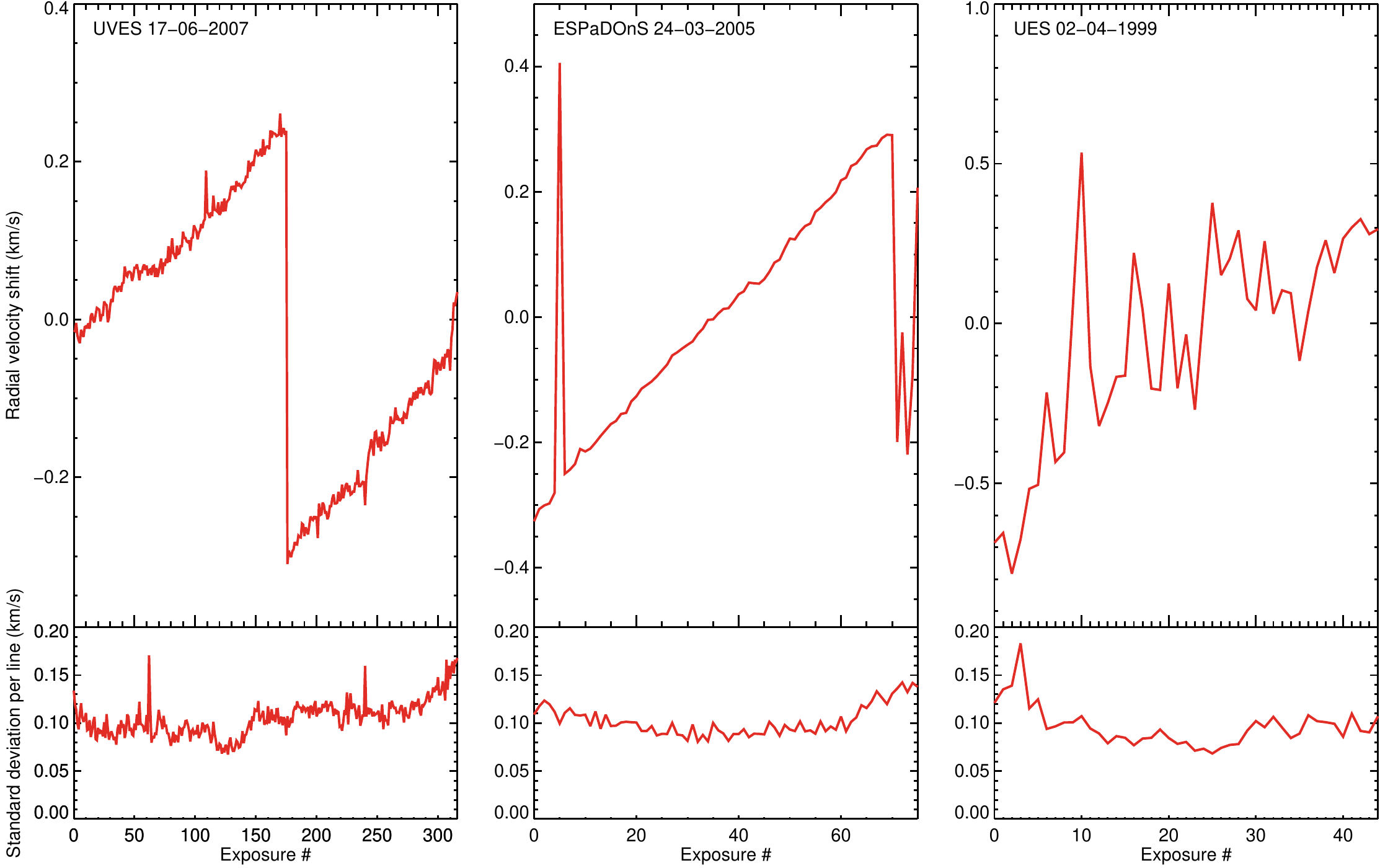}
   \caption{Alignment of the wavelength solution for a single UVES, ESPaDOnS and UES dataset. The top panels indicate the mean shift required to align the absorption lines to their average position. The bottom panels show the residual scatter in the line positions. On average, absorption lines are aligned to within $0.1\kms$ accuracy. The wavelength solution of the UES data is less stable compared to those of UVES and ESPaDOnS, however our procedure is able to align the UES spectra to the same level of accuracy as the other instruments.}%
\label{fig:Alignment}
\end{figure*}

\item[2] \textbf{Wavelength solution:} We adopted the pipeline wavelength-solutions that are provided with the UVES, ESPaDOnS and HARPS-N data and shift these to the rest-frame of the star. In previous work we matched a model stellar template to the lines that we identified in step 1 \citep{Hoeijmakers2015}. However because $\tau$ Boo is a fast-rotating F-star, this matching does not result in a more accurate solution than the pipeline solutions for these datasets. To determine the rest-frame velocity of the star, we first obtained the time-averaged spectrum from the mean flux of each spectral pixel (this is also used in steps 6 and 7). This yields the highest S/N measurement of the stellar spectrum, which is subsequently cross-correlated with a model stellar template that was broadened to the rotation velocity of the star (Section \ref{sec:models}). The resulting cross-correlation strengths are close to 1.0, confirming the quality of the pipeline wavelength solutions (see Fig. \ref{fig:Efficiency}).
\item[3]\textbf{Model injection:} We duplicate each spectral order and inject a model of the planet reflection spectrum (see Section \ref{sec:models}) into the duplicate after appropriate broadening, Doppler-shifting and scaling:
\begin{enumerate}
\item \textbf{Broadening:} The model spectrum is convolved with a Gaussian kernel with a FWHM of $2.61\kms$ to $5.66\kms$ to match the spectral resolution of the respective instrument. It is then blurred with a box-kernel with a width equal to the velocity shift of the planet during the exposure. This is calculated by multiplying the first derivative of the radial velocity at time $t$ with the exposure time of the spectrum. The spectrum is also convolved with a rotation profile\footnote{The rotation kernel is calculated using the IDL routing LSF\_ROTATE with zero limb-darkening applied, with a rotation velocity of $v \sin (i) = \frac{2\pi R_p}{P} \sin (i) = 1.24 \kms$, assuming a tidally locked planet with a radius of $R_p = 1.15R_J$ (also see Sec. \ref{sec:ResultsAlbedo}).}
\item \textbf{Doppler-shifting:} From the known ephemeris and orbital parameters (Table \ref{tab:systemparameters}) we calculate the orbital phase $\varphi$ and the radial velocity of the planet at the time of each observation. The model is then interpolated to the wavelength grid of the data.
\item \textbf{Scaling:} The template is subsequently multiplied by $\epsilon(\lambda)\Phi(\lambda)$. We assume that $\epsilon$ and $\Phi$ are independent of wavelength (grey) and we follow previous authors in adopting a Lambertian phase-function,
\begin{equation}
	\Phi=\frac{\sin{\alpha}+(\pi-\alpha)\cos{\alpha}}{\pi},
\end{equation}

where $\epsilon$ is set to $2.0\times 10^{-5}$ (see Figures \ref{fig:phasecurve} and \ref{fig:toymodel}). The injected and non-injected data sets are treated identically from this step onward. This simultaneous treatment of the model-injected data allows us to quantify the extent to which the data analysis procedure may influence the planet reflection spectrum, and to optimize the parameters of the procedure to maximize the strength at which the injected planet spectrum is retrieved.
 \begin{figure}
   \centering
   \includegraphics[width=1.0\linewidth]{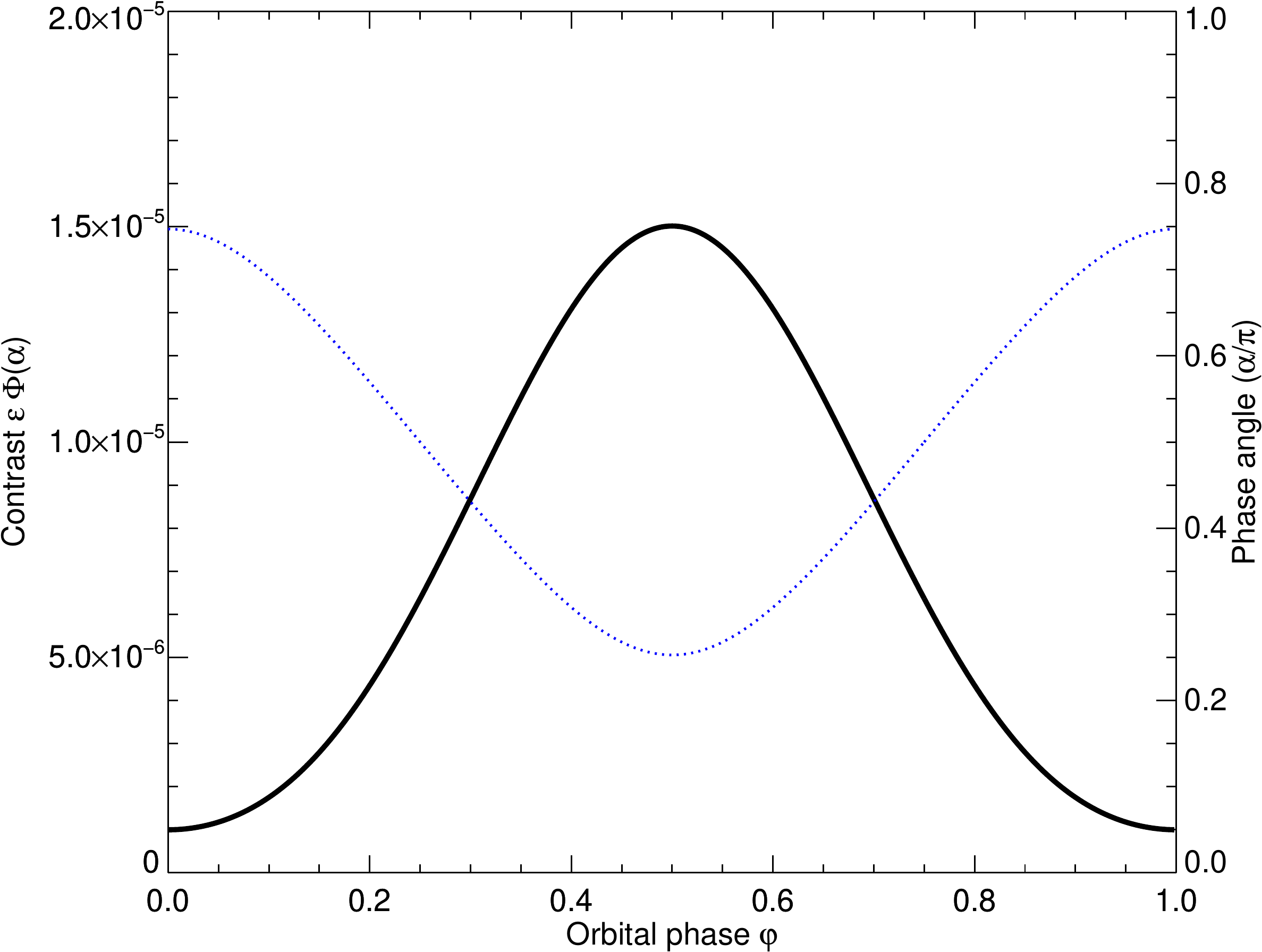}
   \caption{The phase-angle $\alpha(\varphi)$ as function of orbital phase $\varphi$ for an orbital inclination of 44.5$\degrees$(dashed line), and the resulting contrast curve assuming a Lambertian phase function $\Phi(\alpha)$ (solid line). The orbital phase is 0 when the planet is located closest to the observer (i.e. during transit for a planet with an orbital inclination of $\sim 90\degrees$). At this moment, the phase angle is maximal and the contrast is minimal because a small part of the day-side of the planet is in view.}%
\label{fig:phasecurve}
\end{figure}

 \begin{figure}
   \centering
   \includegraphics[width=1.0\linewidth]{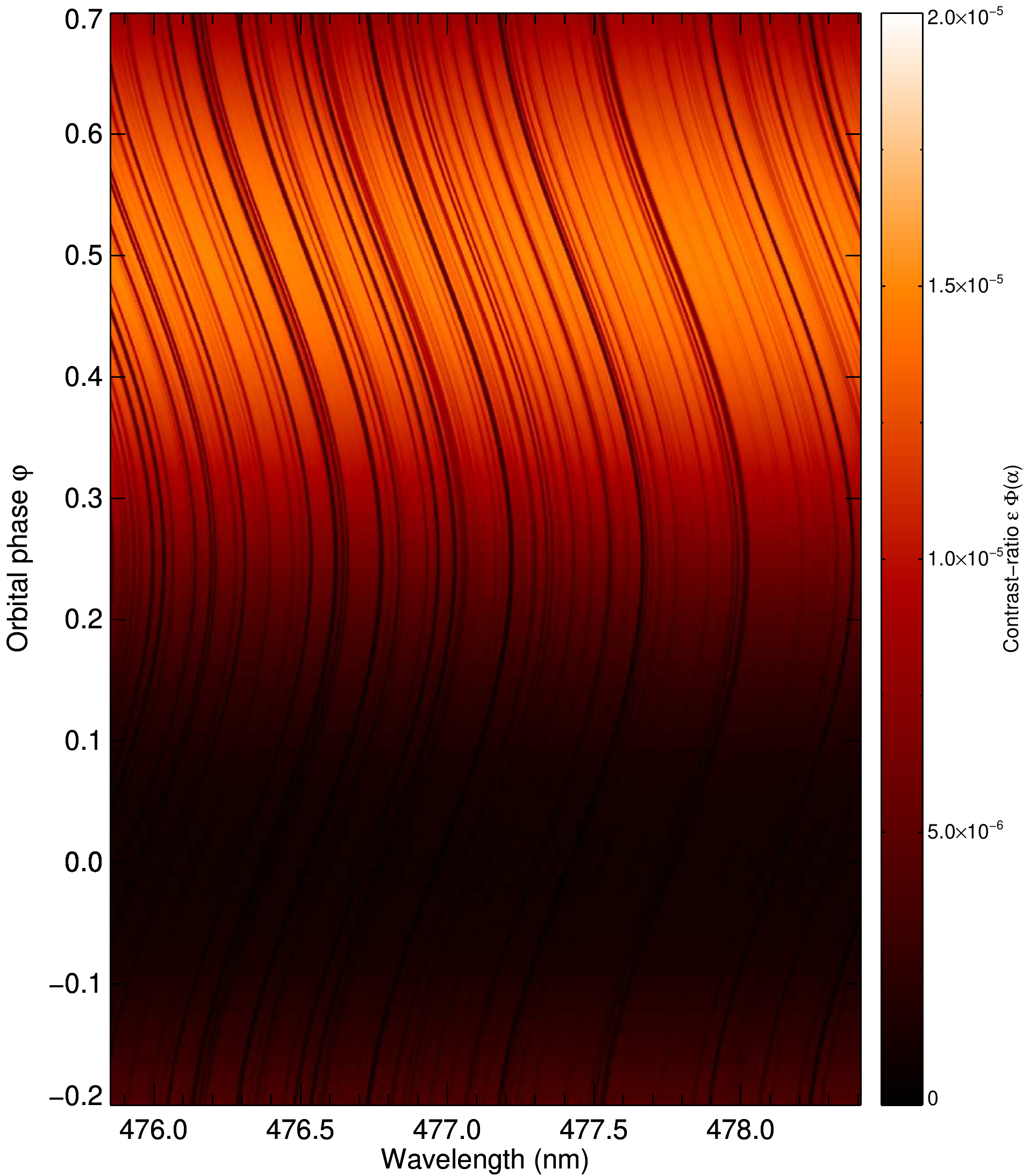}
   \caption{Example of the injected planet model spectra over a complete orbit, broadened to the resolution of UVES and shifted to the instantaneous radial velocity of the planet. The model spectrum is scaled to a level of $\epsilon = 2\times10^{-5}$, but the maximum contrast is less than that because the peak of the phase function $\Phi(\alpha)$ is less than 1.0.}%
\label{fig:toymodel}
\end{figure}

\end{enumerate}

\item[4] \textbf{Removal of the stellar spectrum:} We remove the stellar spectrum by subtracting the time-averaged spectrum from each exposure, as was done in previous work \citep[e.g.][]{Charbonneau1999,Colliercameron2000,Rodler2010}, by subtracting the mean from each column in Fig. \ref{fig:postproc}. This removes the time-constant components of the stellar spectrum, but leaves residual variations - for example those caused by changes in the observing conditions and/or intrinsic stellar variations (panel 2 of Fig. \ref{fig:postproc}).

\item[5]\textbf{Removal of time-dependent residuals:}
For each spectrum we remove broad-band variations by applying a high-pass box-filter. Subsequently, we remove up to 12 principal components to eliminate variations in telluric lines and stellar spectral line shapes. This procedure partly suppresses the planet's spectrum as well. This blind approach of cleaning the data is therefore a trade-off between effective removal of systematics and preserving the planet's reflection spectrum. To optimize the analysis, we perform a grid-search by varying the width of the box-filter and the number of principal components to remove, and choose the combination for which the signal-to-noise of the injected planet signal is maximized after cross-correlation (step 7). Finally, pixel values that are deviant by more than $5\sigma$ from the mean of their column and the two adjacent columns, are set to the mean.

\item[6] \textbf{Normalization by signal to noise:} Stellar absorption lines and low flux levels at the edges of the blaze orders of the spectrograph cause the signal-to-noise to vary within orders. We weigh down noisy wavelengths by dividing each column to its standard deviation, as was also done in for example \citet{Snellen2010}.

\item[7] \textbf{Cross correlation:} We cross-correlate the residuals with a template spectrum that was also used to inject the planet signal in the data in step 3 (see Section \ref{sec:models}). The cross-correlation function (CCF) computed over a range of radial-velocities from $-1600$ to $+1600\kms$ in steps of $1\kms$. For each spectrum, this yields 3201 cross-correlation coefficients, which are expected to peak where the template is shifted to the correct radial velocity of the planet at the time on which the spectrum was obtained. At this velocity, all absorption lines in the planet's spectrum are effectively co-added, causing an enhancement in the cross-correlation-function. The cross-correlation is performed over $\pm1600$ steps in radial velocity to obtain a statistical sample of cross-correlation coefficients over which the random noise level can be measured reliably.

\item[8] \textbf{Masking stellar residuals:} At the rest-frame velocity of the star, the CCFs show residuals that arise from time variability in the stellar spectrum due to for example activity, star-spots and instabilities in the observing conditions and the instruments. These residuals take the form of a vertical structure at the rest-frame velocity of the star in the two-dimensional CCF (see Fig. \ref{fig:CCF}). We mask out all cross-correlation coefficients within $\pm 35\kms$ of the rest-frame velocity. This removes the planet during parts of the orbit where it has a low radial velocity (i.e. when it is in full view at $\varphi \sim 0.5 $), but ensures that only exposures in which the planet's spectrum is Doppler-shifted away from the stellar line-wings are considered in the co-addition of the data.

	\end{enumerate}	

At this stage, a CCF is associated with each spectrum, both with and without injected planet signals.  For each dataset we therefore have $2 \times N_{\textrm{orders}} \times N_{\textrm{exposures}}$ CCFs. For each spectrum, we measure the strength at which the injected template is retrieved by taking the difference between the injected and non-injected CCFs, and dividing with the standard deviation of the non-injected CCF. We call this quantity the 'retrievability'. To combine all of the data, the CCFs are shifted to the rest-frame velocity of the planet, weighed by their respective retrievability and summed.

This weighing scheme ensures that spectra are weighed to account for the phase function $\Phi(\alpha)$ (spectra taken near inferior conjunction will have a lower retrievability because the template spectra was injected at a lower level compared to spectra near superior conjunction), by the number of stellar spectral lines in each order (the cross-correlation is more effective at wavelengths where there are many narrow spectral lines), by all noise sources that degrade the efficiency by which the planet's spectrum can be retrieved and by differences between datasets in terms of observing conditions and data quality. Approximately $5\%$ of all spectral orders are found not to contribute positively to the retrieved signal-to-noise of the injected signal, mostly due to the presence of strong tellurics and CCD artifacts in these spectra. These orders are discarded when co-adding the CCFs.

\begin{figure}
   \centering
   \includegraphics[width=0.9\linewidth]{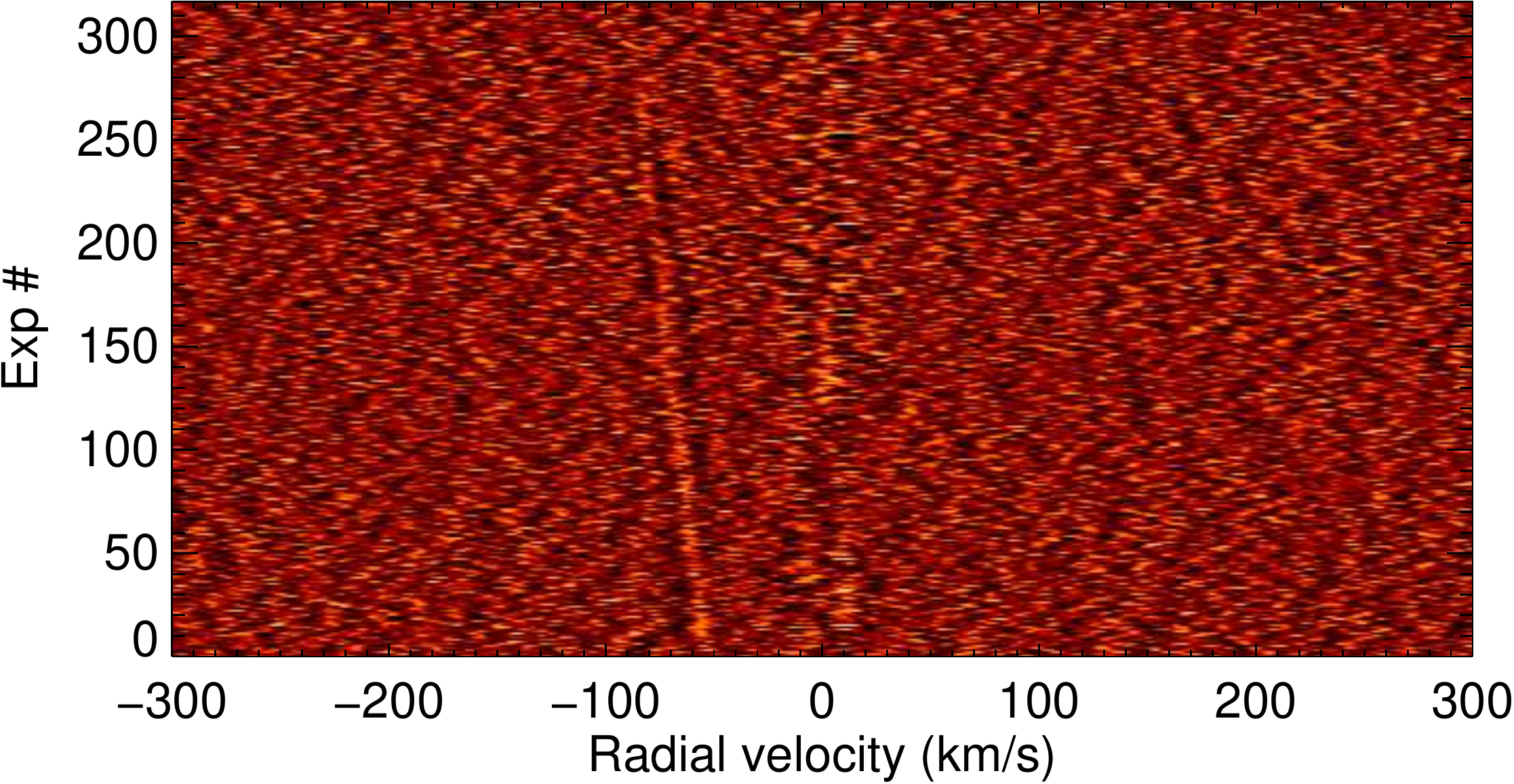}
   \includegraphics[width=0.9\linewidth]{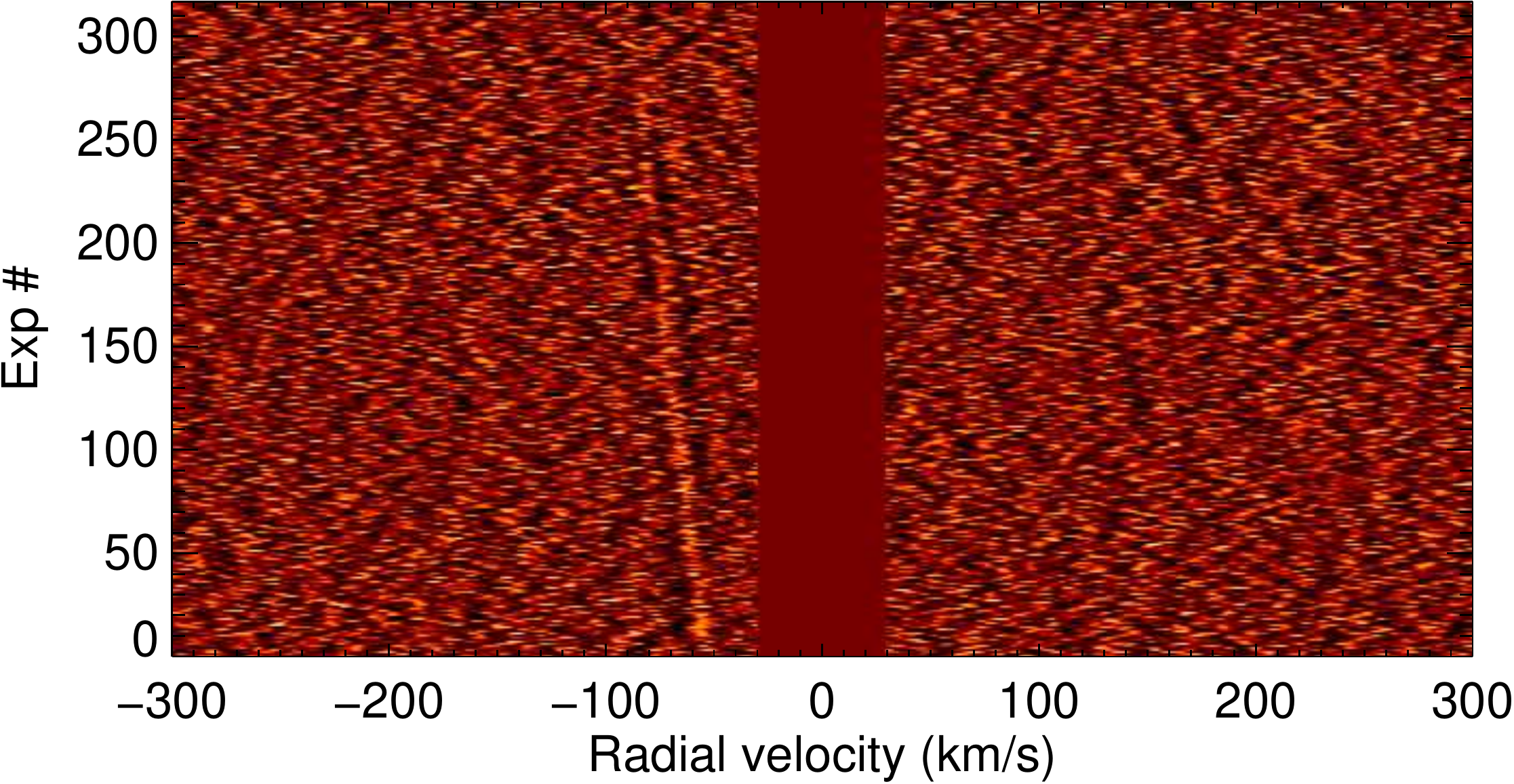}
   \caption{\textbf{Upper Panel:} The combined cross-correlation function of a spectral order of the second UVES night. For illustrative purposes, the planet's spectrum was injected into these data at $\epsilon = 1\times10^{-3}$ to show the slanted cross-correlation peak around $-80 \kms$ due to the changing radial velocity of the planet during the observations. The vertical structure at 0 km/s is caused by residual correlation with remnants of the stellar spectrum. \textbf{Lower Panel:} The same data but with the residual stellar structure at $0 \kms$ masked out. This will prevent it from contaminating non-zero radial-velocities when co-adding the individual cross-correlations at the rest-frame of the planet. }%
\label{fig:CCF}
\end{figure}

\subsection{Template spectra}\label{sec:models}

Although the stellar spectrum is present in the data at high signal-to-noise, it is strongly rotationally broadened due to the short rotation period of the star that is locked to the orbital period of the planet. Therefore, the reflected planet spectrum is expected to exhibit significantly more narrow spectral features making the broadened stellar spectrum a poor model, as already noted by previous authors \citep{Charbonneau1999,Colliercameron2000,Leigh2003,Rodler2010,Rodler2013b}. Therefore, we use a high spectral-resolution stellar photosphere model from the G\"ottingen Spectral Library generated with the PHOENIX radiative transfer code \citep{Husser2013}. The library\footnote{http://phoenix.astro.physik.uni-goettingen.de/?page\_id=15} was queried with $T_{\textrm{eff}}=6300\Kelvin$, $\log g = +4.50$, $[\textrm{Fe}/\textrm{H}]=+0.5$ and $[\alpha/\textrm{M}]=0.0$, to match the literature values for these parameters (Table \ref{tab:systemparameters}).

\begin{figure*}
   \centering
   \includegraphics[width=0.7\linewidth]{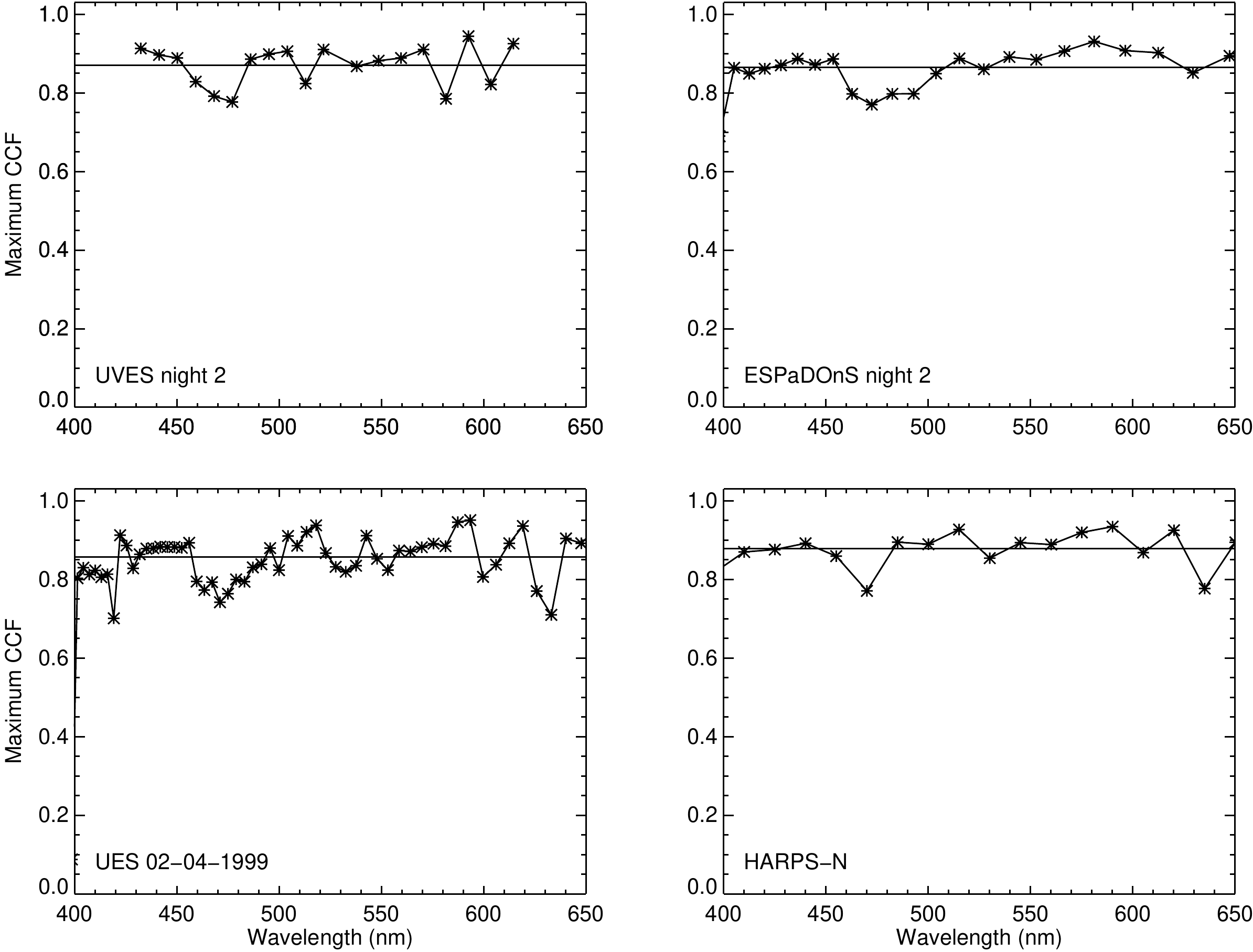}
   \caption{Peak cross-correlation of the time-averaged spectrum of each order with the PHOENIX template spectrum broadened to the rotation velocity of the star. In the four datasets shown, the peak cross-correlation strength averages between 0.85 and 0.90, indicating that the template retrieves the stellar spectrum at high efficiency, and therefore is an accurate model of the planet's reflection spectrum. The cross-correlation is only performed at wavelengths between 450 nm to 650 nm, because the disappearance of the continuum at short wavelengths (hence a break-down of the continuum-normalization needed to perform this cross-correlation) and the presence of tellurics at longer wavelengths cause the correlation of the template with the stellar spectrum to be degraded.
}%
\label{fig:Efficiency}
\end{figure*}

We tested the quality of this model by broadening it to the instrumental resolution and to the rotational velocity of the star, and subsequently cross-correlated it with the time-average spectrum of each spectral order. Examples are shown in Fig. \ref{fig:Efficiency}, demonstrating that mismatches between the template and the observed stellar spectrum on average degrade the cross-correlation by only $13\%$. The model template we use is therefore expected to retrieve the planet's reflection spectrum with $87\%$ efficiency, and we apply this as a correction factor on the end result (Section \ref{sec:ResCCF}).

We measured the gain of using an unbroadened template to be a factor $\sim2.0$ in cross-correlation signal-to-noise at spectral resolutions of $55,000 - 100,000$, compared to a template broadened to $v\sin (i) = 15\kms$.

\begin{figure*}
   \centering
   \includegraphics[width=0.85\linewidth]{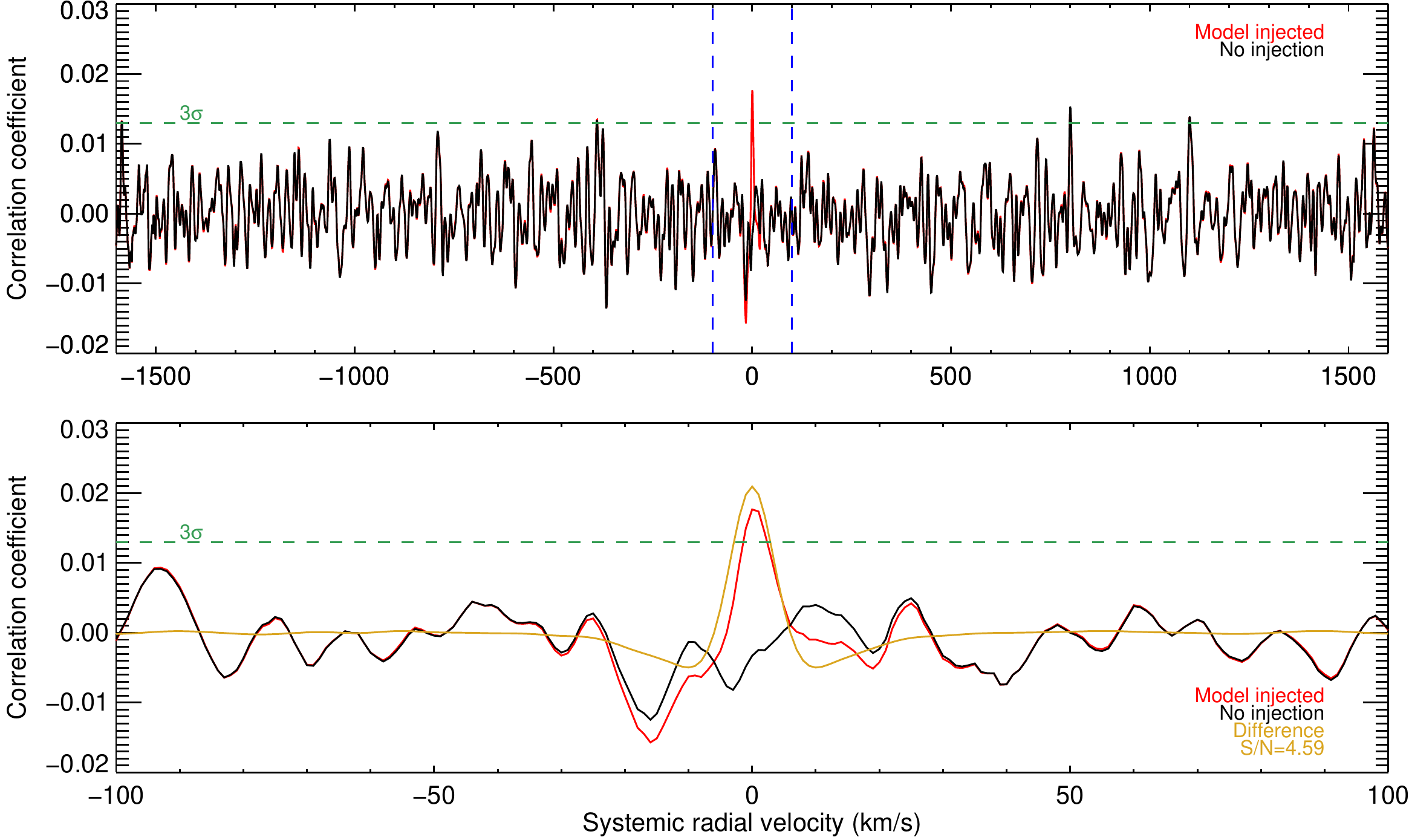}
   \caption{The 1D cross-correlation function after stacking all exposures in all orders of all datasets. The top panel shows the entire cross-correlation function between $\pm 1600 \kms$, while the bottom panel is a zoom-in around $\pm 100 \kms$ for clarity. The red and black lines represent the injected and non-injected data respectively. The gold line is the difference between the two, and the dashed horizontal line is 3$\sigma$ away from the mean of cross-correlation (which is around zero). The model was injected at a strength of $2\times10^{-5}\Phi(\alpha)$, and is retrieved at a level of $\nsig \sigma$. The corresponding $3\sigma$ upper limit of $\epsilon$ is \epslimit, taking into account a factor 0.87 to correct for the efficiency at which the PHOENIX template correlates with the stellar spectrum (see Section \ref{sec:models}).
}%
\label{fig:Finalstack}
\end{figure*}
\begin{figure}
   \centering
   \includegraphics[width=0.9\linewidth]{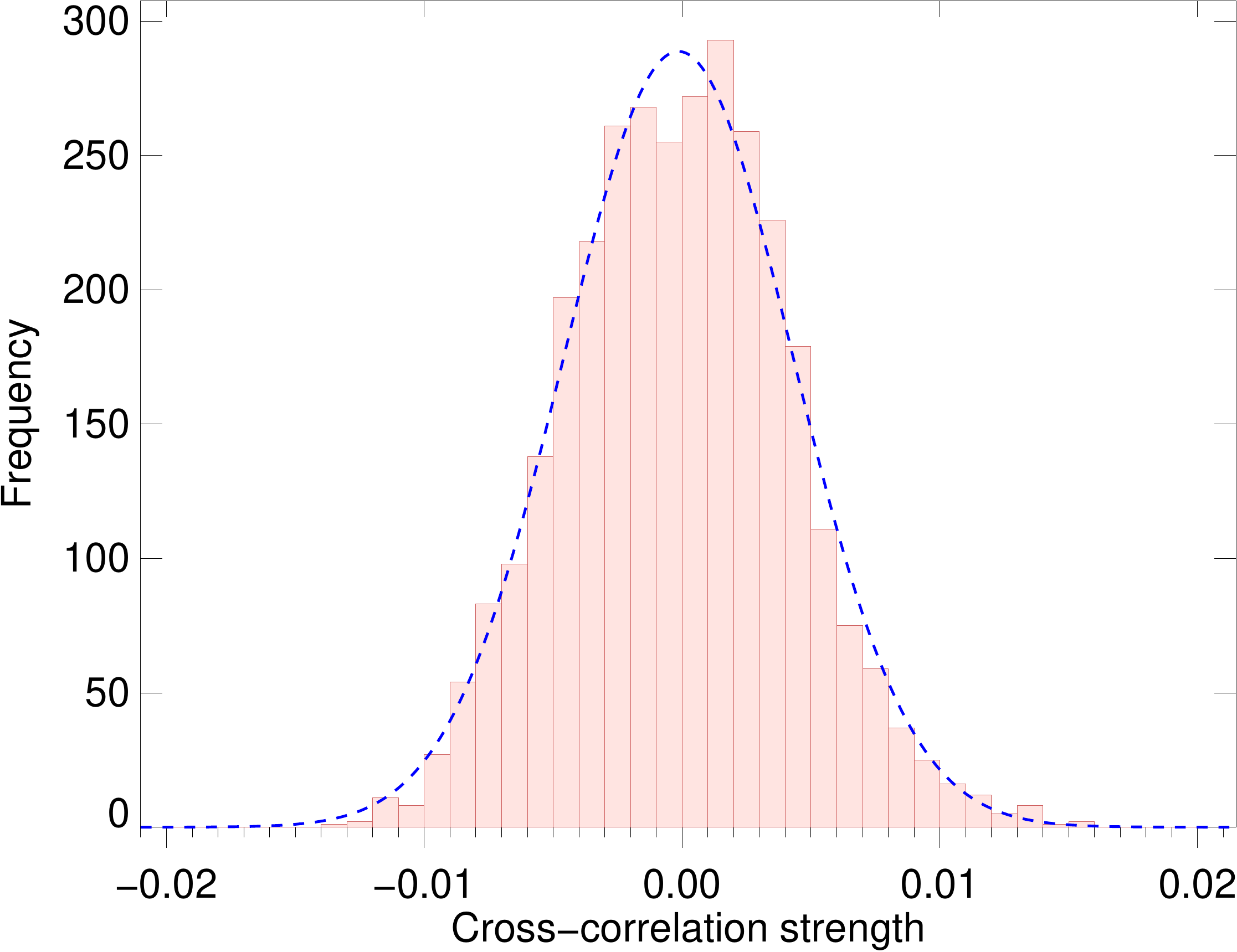}
   \caption{Distribution of the 1D cross-correlation function after stacking all exposures in all orders of all datasets. The blue line is a Gaussian fit to the distribution, of which the standard deviation is used to compute the confidence interval of the injected signal.}%
\label{fig:Finalstack_distribution}
\end{figure}

\section{Results and discussion}\label{sec:ResultsDiscussion}
\subsection{The cross-correlation function}\label{sec:ResCCF}
The final one-dimensional cross-correlation function obtained by combining all the analysed data is shown in Fig. \ref{fig:Finalstack}. We do not detect a significant signal at the rest-frame velocity of the planet. However, the model spectrum injected at a strength of $\epsilon=2.0 \times 10^{-5}$ is retrieved at $\nsig \sigma$ significance. The noise level was measured from the one-dimensional CCF, which was evaluated over $\pm1600$ steps in radial velocity in order to confirm that the noise distribution is Gaussian and that the usage of a $3\sigma$ confidence threshold is prudent (see Fig. \ref{fig:Finalstack_distribution}). As discussed in Section \ref{sec:models}, the PHOENIX template retrieves the rotation-broadened stellar spectrum with an efficiency of $87\%$. Therefore, we establish an upper limit on $\epsilon$ of \epslimit at $3\sigma$ confidence.

\begin{figure}
   \centering
   \includegraphics[width=\linewidth]{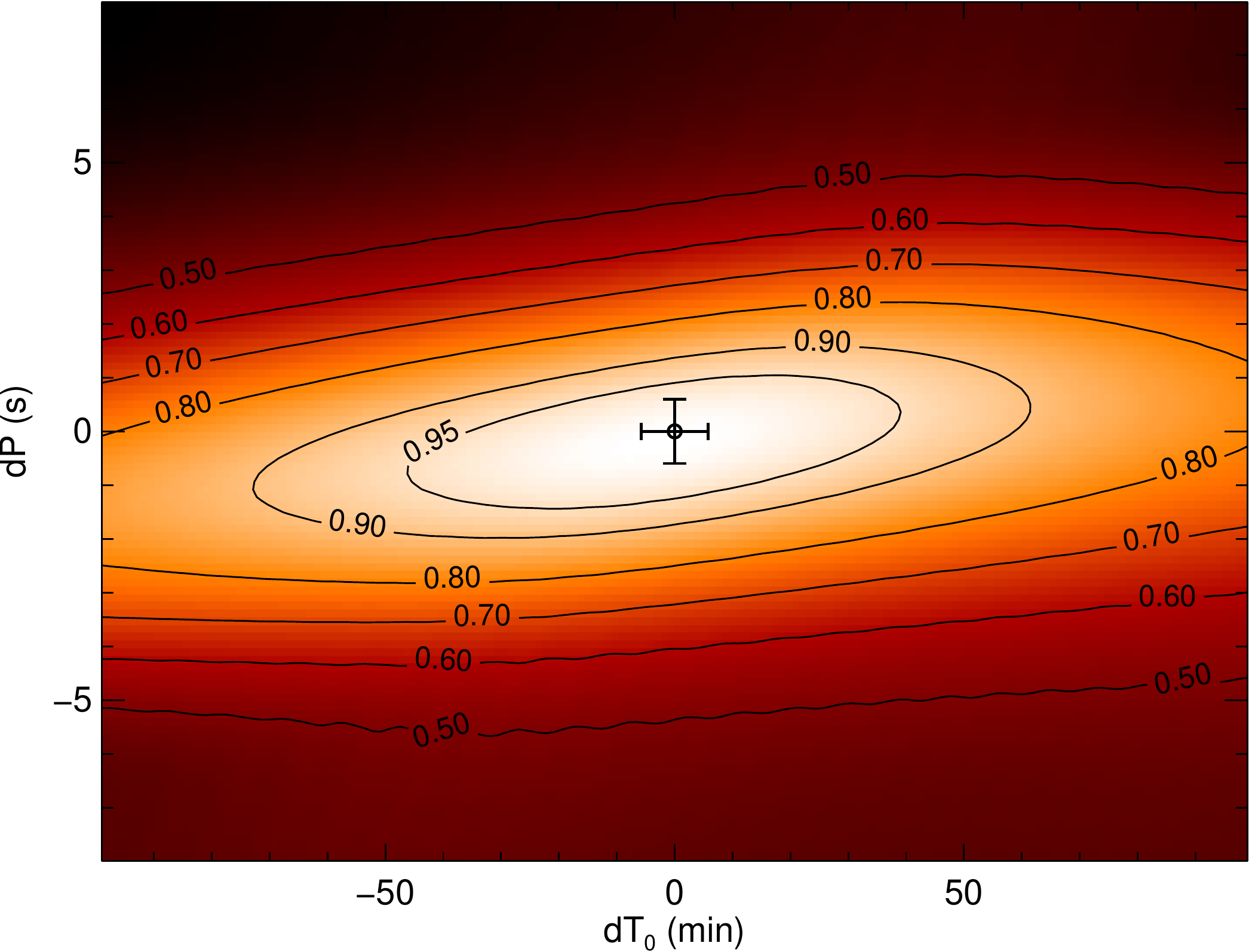}
   \caption{Relative decrease of the retrieved planet signal when stacking the cross-correlation functions using erroneous values for $P$ and $T_0$ in the ephemeris. The retrieved signal is especially sensitive to an error in the orbital period, because the data is obtained over a 15 year timespan. The error bars indicate the known statistical errors on $P$ and $T_0$ as reported by \citet{Brogi2012} and \citet{Borsa2015}. We stack the data at orbital periods that are deviant from the reported period by \citet{Borsa2015} by 15, ten and five times their reported standard error, but find but find no signature of $\tau$ Boo b. This is done to take into account the possibility of a systematic error in the orbital period, which could have affected our analysis.}%
\label{fig:SN_Landscape}
\end{figure}

This result depends on the assumed values of the orbital period $P$ and phase zero-point $T_0$ which have associated uncertainties. We simulated the influence of uncertainties on the ephemeris by stacking the data at small deviations $dP$ and $dT_0$ from the period and phase at with which it was injected, while treating $K_P$ as a fixed constant. Fig. \ref{fig:SN_Landscape} shows the signal-to-noise of the retrieved signal as a function of $dP$ and $dT_0$, normalized to the maximum signal-to-noise at which the signal is retrieved when $dP=dT_0=0$. The analysis is only mildly sensitive to errors in $T_0$, but could be strongly affected by an error in $P$ because the data was taken over a 15 year timespan, spanning $\sim1500$ orbital periods. We therefore repeated the complete analysis while varying the planet orbital period $P$ within $ \pm 15$ times the standard error reported by \citet{Borsa2015}, to take into account the possibility of a large unknown systematic error on the orbital period. However, the resulting upper limits are the same for all trials of $dP$ and no planet signal is retrieved in any of these cases.

\begin{figure}
   \centering
   \includegraphics[width=0.9\linewidth]{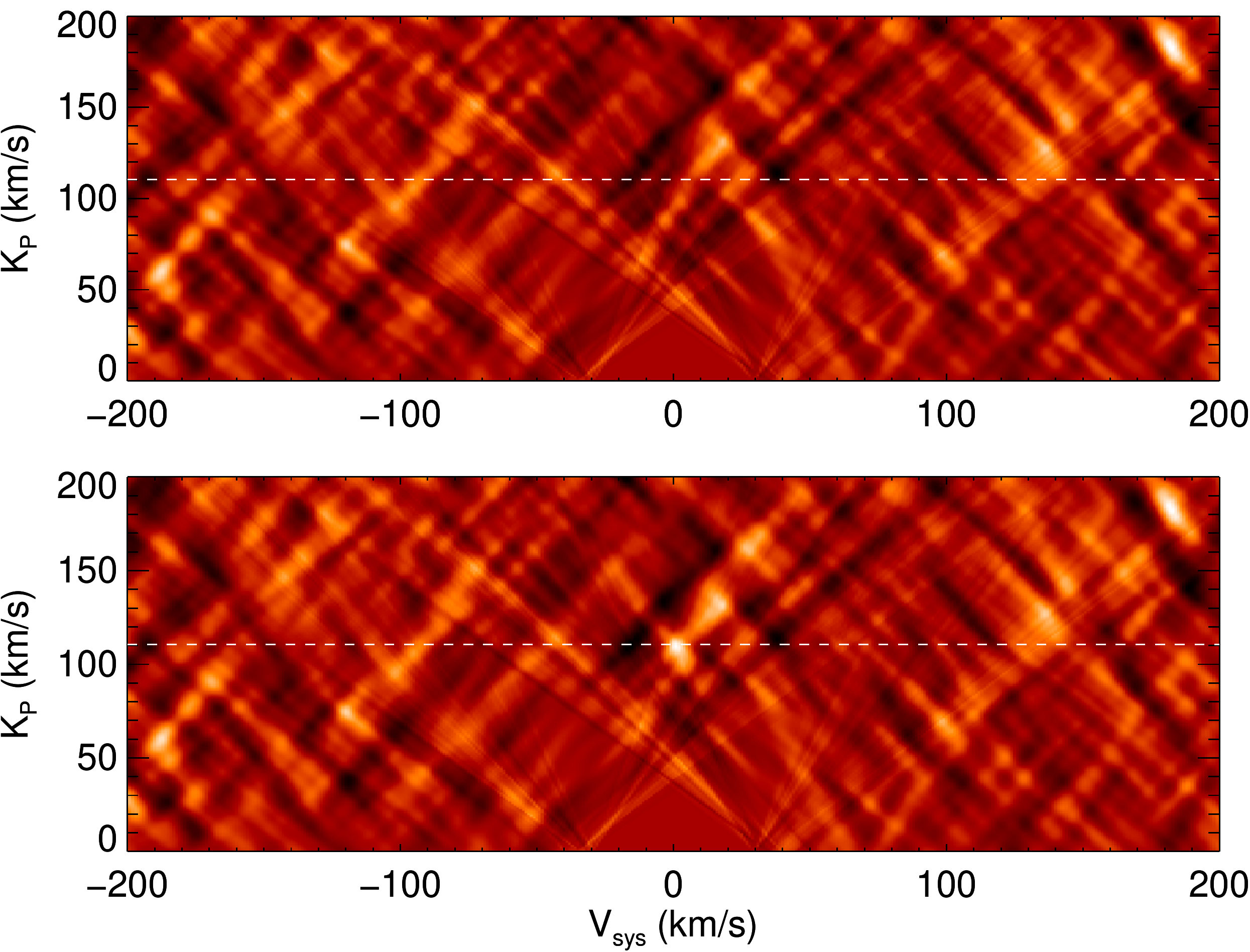}
   \caption{\textbf{Upper panel:} Cross-correlation strength as a function of the rest frame velocity of the system $V_\textrm{sys}$ and the planet radial velocity amplitude $K_p$. \textbf{Lower panel:} The cross-correlation strength for the injected planet signal at $\epsilon = 2\times10^{-5}$. No significant signals are present near the expected $K_p$ of the planet.}
\label{fig:KpVsys}
\end{figure}

To search for a cross-correlation signal at values of $K_p$ other than $110.2\kms$ as determined by \citet{Brogi2012}, we co-added all CCFs for a range of values of $K_p$ (Fig. \ref{fig:KpVsys}). No enhancement in cross-correlation is observed near the expected $v_\textrm{sys}$ and $K_p$ of the planet.

\subsection{The albedo of $\tau$ Boo b}\label{sec:ResultsAlbedo}

\begin{figure*}
   \centering
   \includegraphics[width=0.9\linewidth]{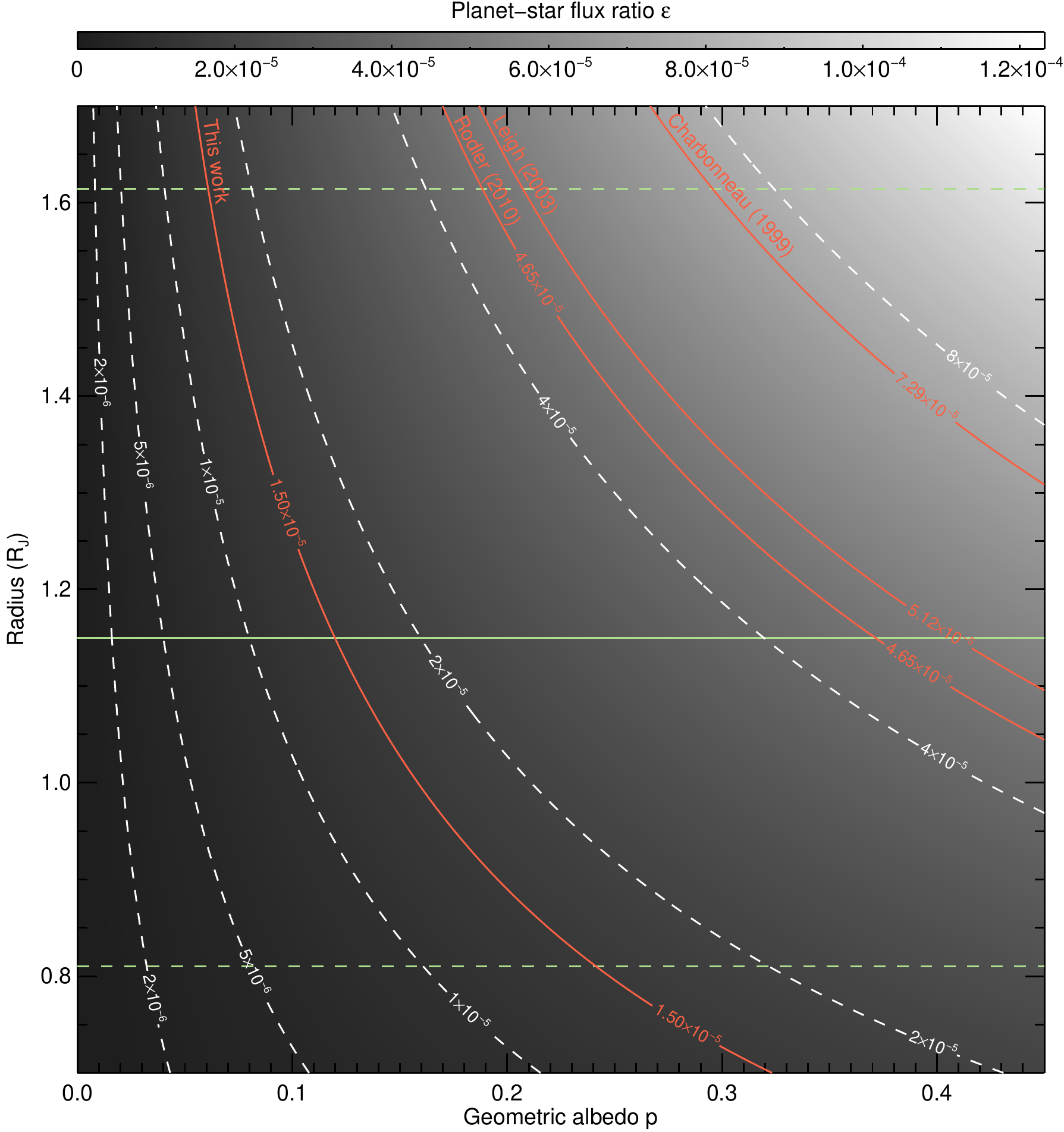}
   \caption{Planet-to-star contrast ratio $\epsilon$ as a function of the geometric albedo $p$ and the planetary radius $R_p$. Solid contour lines indicate the contrast limit of \epslimit \ and the limits reached by \citet{Charbonneau1999}, \citet{Colliercameron2000} and \citet{Rodler2010} converted to 3$\sigma$ confidence. The horizontal lines denote  $R=0.8, 1.15$ and $1.61R_J$, which are the minimum, the mean and the maximum radii of all known transiting hot Jupiters with masses between 3 and $9M_J$.}%
\label{fig:Albedoplot}
\end{figure*}

As follows from Eq. \ref{Eq:Geometric_albedo}, the upper limit on the planet-to-star contrast ratio $\epsilon$ directly constrains the ratio of the projected area of the planet disk ($\sim R_p^2$) and the geometric albedo. Our 3$\sigma$ upper limit of $\epsilon <$\epslimit \ is plotted in Fig. \ref{fig:Albedoplot} as a function of both parameters and compared to the upper limits from \citet{Colliercameron2000}, \citet{Rodler2010} and \citet{Charbonneau1999}, converted to 3$\sigma$ confidence.

Because $\tau$ Boo b is a non-transiting planet, the radius of the planet is unknown and must be estimated from the known population of similar hot Jupiters. We select all transiting planets with masses between $3M_J$ and $9M_J$ and orbital periods less than ten days \citep{Schneider2011} and find a mean radius of $1.15R_J$ for this sample of planets, the same radius as assumed by \citet{Brogi2012}. Combined with the upper limit on $\epsilon$, this places a $3\sigma$ upper limit on the geometric albedo of \plimit.

To date, optical secondary-eclipse observations have been performed of a few dozen transiting hot Jupiters, mostly with the Kepler space observatory \citep[see e.g.][]{Coughlin2012,Esteves2013,Angerhausen2015}.  These have shown that hot Jupiters tend to be dark, with typical visible-light albedo's between 0.06 and 0.11 \citep{Demory2014}, and even cases where the albedo has been shown to be lower than $0.04$ \citep{Kipping2011,Gandolfi2015}.  Our limit of $\epsilon<$\epslimit \ shows that the albedo of $\tau$ Boo b likely lies in a range that is expected for hot Jupiters and that the candidate signals observed by \citet{Leigh2003}, \citet{Rodler2010} and \citet{Rodler2013} are false positives, as correctly hypothesized by these authors.

A low albedo can be explained by strong absorption at visible wavelengths due to the broad wings of alkali absorption lines, Rayleigh scattering by hydrogen and small condensate particles and an absence of a reflective cloud deck \citet{Heng2013}. Because Rayleigh scattering dominates at short wavelengths, the planet is expected to have a blue appearance such as has been observed in HD 18733 b \citep{Evans2013}. We investigated whether such a Rayleigh scattering signal is present in the data by co-adding only the blue spectral orders at wavelengths below 450 nm, roughly matching the band within which \citet{Evans2013} measured the albedo of HD189733 b to be $p=0.32\pm0.15$. Also in this case, we measure no reflection signal from the planet and constrain the average contrast between 380 nm and 450 nm to $3.2\times10^{-5}$.

Using similar methods and observations as used in this work, \citet{Martins2015} claim a detection of 51 Peg b at the level of $\epsilon = 6.0 \pm 0.4$. This requires the planet to have a high albedo of $p$=0.5 for a 1.9$R_J$ planet, or higher if assuming a smaller radius. Their measurement is in stark contrast with the observed trend that hot Jupiters tend to have low albedo's, to which only a few exceptions are known to exist \citep{Demory2011,Shporer2014}. If 51 Peg b indeed has a phase-curve amplitude of $60\pm4$ parts per million, it would be especially suitable for follow-up with the TESS and CHEOPS space telescopes. By measuring the shape of the phase curve, such observations would help to constrain the nature of the scattering particles in the atmosphere of this planet and shed light on the causes for its anomalous brightness \citep{Heng2013,Renyu2015,Oreshenko2016}. 

On the other hand, $\tau$ Boo and other dark hot Jupiters around bright stars will require more sustained observations before their reflected light can be characterized at visible wavelengths. High-resolution ground-based spectroscopy may continue to provide a viable alternative to space-based observations, especially in the absence of the Kepler survey that offered the long observing baselines needed to measure the majority of optical secondary eclipses to date. Notably the ESPRESSO instrument at the VLT will be perfectly suited for these kinds of observations, owing to its high spectral resolution, superior stability and the photon-collecting power of the VLT \citep{Martins2013}.

\section{Conclusions}
Since its discovery, $\tau$ Boo has been observed in various programmes in attempts to detect its reflected light using high dispersion spectroscopy at visible wavelengths \citep{Charbonneau1999,Leigh2003,Rodler2010,Rodler2013b}. These have constrained the planet to be fainter than $5\times10^{-5}$ times the brightness of the star.

In this work, over 2,000 archival UVES, UES, HARPS-N, ESpaDOnS and NARVAL spectra were combined and cross-correlated with a model template to search for the reflected stellar spectrum at the rest-frame velocity of the planet. We are able to rule out planet-to-star contrasts greater than \epslimit \ at $3\sigma$ confidence, under the assumption of a lambertian phase function. The noise level of this analysis is thus \epslimitone, on par with optical secondary eclipse measurements of other hot Jupiters by the Kepler, CoRoT and MOST space observatories. $\tau$ Boo b has a mass of $6.13 \pm 0.17 M_J$, and planets in this mass range have radii of $1.15R_j$ on average. This radius would constrain the geometric albedo to a value of \plimit. A low albedo is consistent with theoretical predictions and a number of studies of other hot Jupiters have shown that such albedo values are indeed common.

\begin{acknowledgements}
      This work is part of the research programmes PEPSci and VICI 639.043.107, both funded by the Dutch Organisation for Scientific Research (NWO), as well as the Leiden Observatory Huygens Fellowship. Snellen acknowledges funding from the European Research Council (ERC) under the European Union’s Horizon 2020 research and innovation programme under grant agreement No 694513. Part of this work is based on observations obtained at the Canada-France-Hawaii Telescope (CFHT) which is operated by the National Research Council of Canada, the Institut National des Sciences de l`Univers of the Centre National de la Recherche Scientique of France, and the University of Hawaii; on observations collected at the European Organisation for Astronomical Research in the Southern Hemisphere under ESO programme 079.C-0413(A); on observations obtained with HARPS-N at the TNG as part of the GAPS programme, and on observations obtained with the UES spectrograph at the WHT. 
\end{acknowledgements}

\bibliographystyle{aa} 
\bibliography{MMP} 

\begin{thebibliography}{80}
\expandafter\ifx\csname natexlab\endcsname\relax\def\natexlab#1{#1}\fi

\bibitem[{{Alonso} {et~al.}(2009{\natexlab{a}}){Alonso}, {Alapini}, {Aigrain},
  {Auvergne}, {Baglin}, {Barbieri}, {Barge}, {Bonomo}, {Bord{\'e}}, {Bouchy},
  {Chaintreuil}, {de La Reza}, {Deeg}, {Deleuil}, {Dvorak}, {Erikson},
  {Fridlund}, {de Oliveira Fialho}, {Gondoin}, {Guillot}, {Hatzes}, {Jorda},
  {Lammer}, {L{\'e}ger}, {Llebaria}, {Magain}, {Mazeh}, {Moutou}, {Ollivier},
  {P{\"a}tzold}, {Pont}, {Queloz}, {Rauer}, {Rouan}, {Schneider}, \&
  {Wuchterl}}]{Alonso2009b}
{Alonso}, R., {Alapini}, A., {Aigrain}, S., {et~al.} 2009{\natexlab{a}}, \aap,
  506, 353

\bibitem[{{Alonso} {et~al.}(2009{\natexlab{b}}){Alonso}, {Guillot}, {Mazeh},
  {Aigrain}, {Alapini}, {Barge}, {Hatzes}, \& {Pont}}]{Alonso2009a}
{Alonso}, R., {Guillot}, T., {Mazeh}, T., {et~al.} 2009{\natexlab{b}}, \aap,
  501, L23

\bibitem[{{Angerhausen} {et~al.}(2015){Angerhausen}, {DeLarme}, \&
  {Morse}}]{Angerhausen2015}
{Angerhausen}, D., {DeLarme}, E., \& {Morse}, J.~A. 2015, \pasp, 127, 1113

\bibitem[{{Atreya} \& {Romani}(1985)}]{Atreya1985}
{Atreya}, S.~K. \& {Romani}, P.~N. 1985, {Photochemistry and clouds of Jupiter,
  Saturn and Uranus}, 17--68

\bibitem[{{Auri{\`e}re}(2003)}]{Auriere2003}
{Auri{\`e}re}, M. 2003, in EAS Publications Series, Vol.~9, EAS Publications
  Series, ed. J.~{Arnaud} \& N.~{Meunier}, 105

\bibitem[{{Baliunas} {et~al.}(1997){Baliunas}, {Henry}, {Donahue}, {Fekel}, \&
  {Soon}}]{Baliunas1997}
{Baliunas}, S.~L., {Henry}, G.~W., {Donahue}, R.~A., {Fekel}, F.~C., \& {Soon},
  W.~H. 1997, \apjl, 474, L119

\bibitem[{{Borsa} {et~al.}(2015){Borsa}, {Scandariato}, {Rainer}, {Bignamini},
  {Maggio}, {Poretti}, {Lanza}, {Di Mauro}, {Benatti}, {Biazzo}, {Bonomo},
  {Damasso}, {Esposito}, {Gratton}, {Affer}, {Barbieri}, {Boccato}, {Claudi},
  {Cosentino}, {Covino}, {Desidera}, {Fiorenzano}, {Gandolfi}, {Harutyunyan},
  {Maldonado}, {Micela}, {Molaro}, {Molinari}, {Pagano}, {Pillitteri},
  {Piotto}, {Shkolnik}, {Silvotti}, {Smareglia}, {Southworth}, {Sozzetti}, \&
  {Stelzer}}]{Borsa2015}
{Borsa}, F., {Scandariato}, G., {Rainer}, M., {et~al.} 2015, \aap, 578, A64

\bibitem[{{Brogi} {et~al.}(2012){Brogi}, {Snellen}, {de Kok}, {Albrecht},
  {Birkby}, \& {de Mooij}}]{Brogi2012}
{Brogi}, M., {Snellen}, I.~A.~G., {de Kok}, R.~J., {et~al.} 2012, \nat, 486,
  502

\bibitem[{{Brogi} {et~al.}(2013){Brogi}, {Snellen}, {de Kok}, {Albrecht},
  {Birkby}, \& {de Mooij}}]{Brogi2013}
{Brogi}, M., {Snellen}, I.~A.~G., {de Kok}, R.~J., {et~al.} 2013, \apj, 767, 27

\bibitem[{{Burrows} {et~al.}(2008){Burrows}, {Budaj}, \&
  {Hubeny}}]{Burrows2008}
{Burrows}, A., {Budaj}, J., \& {Hubeny}, I. 2008, \apj, 678, 1436

\bibitem[{{Butler} {et~al.}(1997){Butler}, {Marcy}, {Williams}, {Hauser}, \&
  {Shirts}}]{Butler1997}
{Butler}, R.~P., {Marcy}, G.~W., {Williams}, E., {Hauser}, H., \& {Shirts}, P.
  1997, \apjl, 474, L115

\bibitem[{Charbonneau {et~al.}(1999)Charbonneau, Noyes, Korzennik, Nisenson,
  Jha, Vogt, \& Kibrick}]{Charbonneau1999}
Charbonneau, D., Noyes, R.~W., Korzennik, S.~G., {et~al.} 1999, The
  Astrophysical Journal Letters, 522, L145

\bibitem[{{Christiansen} {et~al.}(2010){Christiansen}, {Ballard},
  {Charbonneau}, {Madhusudhan}, {Seager}, {Holman}, {Wellnitz}, {Deming},
  {A'Hearn}, \& {EPOXI Team}}]{Christiansen2009}
{Christiansen}, J.~L., {Ballard}, S., {Charbonneau}, D., {et~al.} 2010, \apj,
  710, 97

\bibitem[{{Collier Cameron} {et~al.}(2000){Collier Cameron}, {Horne}, {James},
  {Penny}, \& {Semel}}]{Colliercameron2000}
{Collier Cameron}, A., {Horne}, K., {James}, D., {Penny}, A., \& {Semel}, M.
  2000, ArXiv Astrophysics e-prints [\eprint{astro-ph/0012186}]

\bibitem[{{Collier Cameron} {et~al.}(1999){Collier Cameron}, {Horne}, {Penny},
  \& {James}}]{Colliercameron1999}
{Collier Cameron}, A., {Horne}, K., {Penny}, A., \& {James}, D. 1999, \nat,
  402, 751

\bibitem[{{Collier Cameron} {et~al.}(2002){Collier Cameron}, {Horne}, {Penny},
  \& {Leigh}}]{Colliercameron2002}
{Collier Cameron}, A., {Horne}, K., {Penny}, A., \& {Leigh}, C. 2002, \mnras,
  330, 187

\bibitem[{{Cosentino} {et~al.}(2012){Cosentino}, {Lovis}, {Pepe}, {Collier
  Cameron}, {Latham}, {Molinari}, {Udry}, {Bezawada}, {Black}, {Born},
  {Buchschacher}, {Charbonneau}, {Figueira}, {Fleury}, {Galli}, {Gallie},
  {Gao}, {Ghedina}, {Gonzalez}, {Gonzalez}, {Guerra}, {Henry}, {Horne},
  {Hughes}, {Kelly}, {Lodi}, {Lunney}, {Maire}, {Mayor}, {Micela}, {Ordway},
  {Peacock}, {Phillips}, {Piotto}, {Pollacco}, {Queloz}, {Rice}, {Riverol},
  {Riverol}, {San Juan}, {Sasselov}, {Segransan}, {Sozzetti}, {Sosnowska},
  {Stobie}, {Szentgyorgyi}, {Vick}, \& {Weber}}]{Cosentino2012}
{Cosentino}, R., {Lovis}, C., {Pepe}, F., {et~al.} 2012, in \procspie, Vol.
  8446, Ground-based and Airborne Instrumentation for Astronomy IV, 84461V

\bibitem[{{Coughlin} \& {L{\'o}pez-Morales}(2012)}]{Coughlin2012}
{Coughlin}, J.~L. \& {L{\'o}pez-Morales}, M. 2012, \aj, 143, 39

\bibitem[{{Covino} {et~al.}(2013){Covino}, {Esposito}, {Barbieri}, {Mancini},
  {Nascimbeni}, {Claudi}, {Desidera}, {Gratton}, {Lanza}, {Sozzetti}, {Biazzo},
  {Affer}, {Gandolfi}, {Munari}, {Pagano}, {Bonomo}, {Collier Cameron},
  {H{\'e}brard}, {Maggio}, {Messina}, {Micela}, {Molinari}, {Pepe}, {Piotto},
  {Ribas}, {Santos}, {Southworth}, {Shkolnik}, {Triaud}, {Bedin}, {Benatti},
  {Boccato}, {Bonavita}, {Borsa}, {Borsato}, {Brown}, {Carolo}, {Ciceri},
  {Cosentino}, {Damasso}, {Faedi}, {Mart{\'{\i}}nez Fiorenzano}, {Latham},
  {Lovis}, {Mordasini}, {Nikolov}, {Poretti}, {Rainer}, {Rebolo L{\'o}pez},
  {Scandariato}, {Silvotti}, {Smareglia}, {Alcal{\'a}}, {Cunial}, {Di
  Fabrizio}, {Di Mauro}, {Giacobbe}, {Granata}, {Harutyunyan}, {Knapic},
  {Lattanzi}, {Leto}, {Lodato}, {Malavolta}, {Marzari}, {Molinaro},
  {Nardiello}, {Pedani}, {Prisinzano}, \& {Turrini}}]{Covino2013}
{Covino}, E., {Esposito}, M., {Barbieri}, M., {et~al.} 2013, \aap, 554, A28

\bibitem[{{Cowan} \& {Agol}(2011)}]{Cowan2011}
{Cowan}, N.~B. \& {Agol}, E. 2011, \apj, 729, 54

\bibitem[{{Czesla} {et~al.}(2015){Czesla}, {Klocov{\'a}}, {Khalafinejad},
  {Wolter}, \& {Schmitt}}]{Czesla2015}
{Czesla}, S., {Klocov{\'a}}, T., {Khalafinejad}, S., {Wolter}, U., \&
  {Schmitt}, J.~H.~M.~M. 2015, \aap, 582, A51

\bibitem[{{Dekker} {et~al.}(2000){Dekker}, {D'Odorico}, {Kaufer}, {Delabre}, \&
  {Kotzlowski}}]{Dekker2000}
{Dekker}, H., {D'Odorico}, S., {Kaufer}, A., {Delabre}, B., \& {Kotzlowski}, H.
  2000, in \procspie, Vol. 4008, Optical and IR Telescope Instrumentation and
  Detectors, ed. M.~{Iye} \& A.~F. {Moorwood}, 534--545

\bibitem[{{Demory}(2014)}]{Demory2014}
{Demory}, B.-O. 2014, \apjl, 789, L20

\bibitem[{{Demory} {et~al.}(2011){Demory}, {Seager}, {Madhusudhan}, {Kjeldsen},
  {Christensen-Dalsgaard}, {Gillon}, {Rowe}, {Welsh}, {Adams}, {Dupree},
  {McCarthy}, {Kulesa}, {Borucki}, \& {Koch}}]{Demory2011}
{Demory}, B.-O., {Seager}, S., {Madhusudhan}, N., {et~al.} 2011, \apjl, 735,
  L12

\bibitem[{{D{\'e}sert} {et~al.}(2011){D{\'e}sert}, {Charbonneau}, {Fortney},
  {Madhusudhan}, {Knutson}, {Fressin}, {Deming}, {Borucki}, {Brown},
  {Caldwell}, {Ford}, {Gilliland}, {Latham}, {Marcy}, \& {Seager}}]{Desert2011}
{D{\'e}sert}, J.-M., {Charbonneau}, D., {Fortney}, J.~J., {et~al.} 2011, \apjs,
  197, 11

\bibitem[{{Desidera} {et~al.}(2013){Desidera}, {Sozzetti}, {Bonomo}, {Gratton},
  {Poretti}, {Claudi}, {Latham}, {Affer}, {Cosentino}, {Damasso}, {Esposito},
  {Giacobbe}, {Malavolta}, {Nascimbeni}, {Piotto}, {Rainer}, {Scardia},
  {Schmid}, {Lanza}, {Micela}, {Pagano}, {Bedin}, {Biazzo}, {Borsa}, {Carolo},
  {Covino}, {Faedi}, {H{\'e}brard}, {Lovis}, {Maggio}, {Mancini}, {Marzari},
  {Messina}, {Molinari}, {Munari}, {Pepe}, {Santos}, {Scandariato}, {Shkolnik},
  \& {Southworth}}]{Desidera2013}
{Desidera}, S., {Sozzetti}, A., {Bonomo}, A.~S., {et~al.} 2013, \aap, 554, A29

\bibitem[{{Donati}(2003)}]{Donati2003}
{Donati}, J.-F. 2003, in Astronomical Society of the Pacific Conference Series,
  Vol. 307, Solar Polarization, ed. J.~{Trujillo-Bueno} \& J.~{Sanchez
  Almeida}, 41

\bibitem[{{Donati} {et~al.}(2008){Donati}, {Moutou}, {Far{\`e}s}, {Bohlender},
  {Catala}, {Deleuil}, {Shkolnik}, {Collier Cameron}, {Jardine}, \&
  {Walker}}]{Donati2008b}
{Donati}, J.-F., {Moutou}, C., {Far{\`e}s}, R., {et~al.} 2008, \mnras, 385,
  1179

\bibitem[{{Esteves} {et~al.}(2013){Esteves}, {De Mooij}, \&
  {Jayawardhana}}]{Esteves2013}
{Esteves}, L.~J., {De Mooij}, E.~J.~W., \& {Jayawardhana}, R. 2013, \apj, 772,
  51

\bibitem[{{Evans} {et~al.}(2013){Evans}, {Pont}, {Sing}, {Aigrain}, {Barstow},
  {D{\'e}sert}, {Gibson}, {Heng}, {Knutson}, \& {Lecavelier des
  Etangs}}]{Evans2013}
{Evans}, T.~M., {Pont}, F., {Sing}, D.~K., {et~al.} 2013, \apjl, 772, L16

\bibitem[{{Fares} {et~al.}(2009){Fares}, {Donati}, {Moutou}, {Bohlender},
  {Catala}, {Deleuil}, {Shkolnik}, {Collier Cameron}, {Jardine}, \&
  {Walker}}]{Fares2009}
{Fares}, R., {Donati}, J.-F., {Moutou}, C., {et~al.} 2009, \mnras, 398, 1383

\bibitem[{{Fressin} {et~al.}(2013){Fressin}, {Torres}, {Charbonneau}, {Bryson},
  {Christiansen}, {Dressing}, {Jenkins}, {Walkowicz}, \&
  {Batalha}}]{Fressin2013}
{Fressin}, F., {Torres}, G., {Charbonneau}, D., {et~al.} 2013, \apj, 766, 81

\bibitem[{{Gandolfi} {et~al.}(2015){Gandolfi}, {Parviainen}, {Deeg}, {Lanza},
  {Fridlund}, {Prada Moroni}, {Alonso}, {Augusteijn}, {Cabrera}, {Evans},
  {Geier}, {Hatzes}, {Holczer}, {Hoyer}, {Kangas}, {Mazeh}, {Pagano}, {Tal-Or},
  \& {Tingley}}]{Gandolfi2015}
{Gandolfi}, D., {Parviainen}, H., {Deeg}, H.~J., {et~al.} 2015, \aap, 576, A11

\bibitem[{{Gandolfi} {et~al.}(2013){Gandolfi}, {Parviainen}, {Fridlund},
  {Hatzes}, {Deeg}, {Frasca}, {Lanza}, {Prada Moroni}, {Tognelli}, {McQuillan},
  {Aigrain}, {Alonso}, {Antoci}, {Cabrera}, {Carone}, {Csizmadia}, {Djupvik},
  {Guenther}, {Jessen-Hansen}, {Ofir}, \& {Telting}}]{Gandolfi2013}
{Gandolfi}, D., {Parviainen}, H., {Fridlund}, M., {et~al.} 2013, \aap, 557, A74

\bibitem[{{Hale}(1994)}]{Hale1994}
{Hale}, A. 1994, \aj, 107, 306

\bibitem[{{Haswell}(2010)}]{Haswell2010}
{Haswell}, C.~A. 2010, {Transiting Exoplanets}

\bibitem[{{Heng} \& {Demory}(2013)}]{Heng2013}
{Heng}, K. \& {Demory}, B.-O. 2013, \apj, 777, 100

\bibitem[{{Hilton}(1992)}]{Hilton1992}
{Hilton}, J.~L. 1992, {Explanatory Supplement to the Astronomical Almanac} (20
  Edgehill Road, Mill Valley, CA 94941, USA: {University Science Books}), 383

\bibitem[{{Hoeijmakers} {et~al.}(2015){Hoeijmakers}, {de Kok}, {Snellen},
  {Brogi}, {Birkby}, \& {Schwarz}}]{Hoeijmakers2015}
{Hoeijmakers}, H.~J., {de Kok}, R.~J., {Snellen}, I.~A.~G., {et~al.} 2015,
  \aap, 575, A20

\bibitem[{{Howard} {et~al.}(2012){Howard}, {Marcy}, {Bryson}, {Jenkins},
  {Rowe}, {Batalha}, {Borucki}, {Koch}, {Dunham}, {Gautier}, {Van Cleve},
  {Cochran}, {Latham}, {Lissauer}, {Torres}, {Brown}, {Gilliland}, {Buchhave},
  {Caldwell}, {Christensen-Dalsgaard}, {Ciardi}, {Fressin}, {Haas}, {Howell},
  {Kjeldsen}, {Seager}, {Rogers}, {Sasselov}, {Steffen}, {Basri},
  {Charbonneau}, {Christiansen}, {Clarke}, {Dupree}, {Fabrycky}, {Fischer},
  {Ford}, {Fortney}, {Tarter}, {Girouard}, {Holman}, {Johnson}, {Klaus},
  {Machalek}, {Moorhead}, {Morehead}, {Ragozzine}, {Tenenbaum}, {Twicken},
  {Quinn}, {Isaacson}, {Shporer}, {Lucas}, {Walkowicz}, {Welsh}, {Boss},
  {Devore}, {Gould}, {Smith}, {Morris}, {Prsa}, {Morton}, {Still}, {Thompson},
  {Mullally}, {Endl}, \& {MacQueen}}]{Howard2012}
{Howard}, A.~W., {Marcy}, G.~W., {Bryson}, S.~T., {et~al.} 2012, \apjs, 201, 15

\bibitem[{{Hu} {et~al.}(2015){Hu}, {Demory}, {Seager}, {Lewis}, \&
  {Showman}}]{Renyu2015}
{Hu}, R., {Demory}, B.-O., {Seager}, S., {Lewis}, N., \& {Showman}, A.~P. 2015,
  \apj, 802, 51

\bibitem[{{Husser} {et~al.}(2013){Husser}, {Wende-von Berg}, {Dreizler},
  {Homeier}, {Reiners}, {Barman}, \& {Hauschildt}}]{Husser2013}
{Husser}, T.-O., {Wende-von Berg}, S., {Dreizler}, S., {et~al.} 2013, \aap,
  553, A6

\bibitem[{{Irwin} {et~al.}(2017){Irwin}, {Wong}, {Simon}, {Orton}, \&
  {Toledo}}]{Irwin2017}
{Irwin}, P.~G.~J., {Wong}, M.~H., {Simon}, A.~A., {Orton}, G.~S., \& {Toledo},
  D. 2017, \icarus, 288, 99

\bibitem[{{Jones, A.} {et~al.}(2013){Jones, A.}, {Noll, S.}, {Kausch, W.},
  {Szyszka, C.}, \& {Kimeswenger, S.}}]{Jones2013}
{Jones, A.}, {Noll, S.}, {Kausch, W.}, {Szyszka, C.}, \& {Kimeswenger, S.}
  2013, \aap, 560, A91

\bibitem[{{Kipping} \& {Spiegel}(2011)}]{Kipping2011}
{Kipping}, D.~M. \& {Spiegel}, D.~S. 2011, \mnras, 417, L88

\bibitem[{Leigh {et~al.}(2003a)Leigh, Cameron, Horne, Penny, \&
  James}]{Leigh2003}
Leigh, C., Cameron, A.~C., Horne, K., Penny, A., \& James, D. 2003a, Monthly
  Notices of the Royal Astronomical Society, 344, 1271

\bibitem[{{Leigh} {et~al.}(2003b){Leigh}, {Collier Cameron}, {Udry}, {Donati},
  {Horne}, {James}, \& {Penny}}]{Leigh2003b}
{Leigh}, C., {Collier Cameron}, A., {Udry}, S., {et~al.} 2003b, \mnras, 346,
  L16

\bibitem[{{Marley} {et~al.}(1999){Marley}, {Gelino}, {Stephens}, {Lunine}, \&
  {Freedman}}]{Marley1999}
{Marley}, M.~S., {Gelino}, C., {Stephens}, D., {Lunine}, J.~I., \& {Freedman},
  R. 1999, \apj, 513, 879

\bibitem[{{Martins} {et~al.}(2013){Martins}, {Figueira}, {Santos}, \&
  {Lovis}}]{Martins2013}
{Martins}, J.~H.~C., {Figueira}, P., {Santos}, N.~C., \& {Lovis}, C. 2013,
  \mnras, 436, 1215

\bibitem[{{Martins} {et~al.}(2015){Martins}, {Santos}, {Figueira}, {Faria},
  {Montalto}, {Boisse}, {Ehrenreich}, {Lovis}, {Mayor}, {Melo}, {Pepe},
  {Sousa}, {Udry}, \& {Cunha}}]{Martins2015}
{Martins}, J.~H.~C., {Santos}, N.~C., {Figueira}, P., {et~al.} 2015, \aap, 576,
  A134

\bibitem[{{Mayor} {et~al.}(1995){Mayor}, {Queloz}, {Marcy}, {Butler}, {Noyes},
  {Korzennik}, {Krockenberger}, {Nisenson}, {Brown}, {Kennelly}, {Rowland},
  {Horner}, {Burki}, {Burnet}, \& {Kunzli}}]{Mayor1995}
{Mayor}, M., {Queloz}, D., {Marcy}, G., {et~al.} 1995, \iaucirc, 6251, 1

\bibitem[{{Morris} {et~al.}(2013){Morris}, {Mandell}, \& {Deming}}]{Morris2013}
{Morris}, B.~M., {Mandell}, A.~M., \& {Deming}, D. 2013, \apjl, 764, L22

\bibitem[{{Moses} {et~al.}(1995){Moses}, {Rages}, \& {Pollack}}]{Moses1995}
{Moses}, J.~I., {Rages}, K., \& {Pollack}, J.~B. 1995, \icarus, 113, 232

\bibitem[{{Nidever} {et~al.}(2002){Nidever}, {Marcy}, {Butler}, {Fischer}, \&
  {Vogt}}]{Nidever2002}
{Nidever}, D.~L., {Marcy}, G.~W., {Butler}, R.~P., {Fischer}, D.~A., \& {Vogt},
  S.~S. 2002, \apjs, 141, 503

\bibitem[{{Noll, S.} {et~al.}(2012){Noll, S.}, {Kausch, W.}, {Barden, M.},
  {Jones, A. M.}, {Szyszka, C.}, {Kimeswenger, S.}, \& {Vinther,
  J.}}]{Noll2012}
{Noll, S.}, {Kausch, W.}, {Barden, M.}, {et~al.} 2012, \aap, 543, A92

\bibitem[{{Oreshenko} {et~al.}(2016){Oreshenko}, {Heng}, \&
  {Demory}}]{Oreshenko2016}
{Oreshenko}, M., {Heng}, K., \& {Demory}, B.-O. 2016, \mnras, 457, 3420

\bibitem[{{Patience} {et~al.}(2002){Patience}, {White}, {Ghez}, {McCabe},
  {McLean}, {Larkin}, {Prato}, {Kim}, {Lloyd}, {Liu}, {Graham}, {Macintosh},
  {Gavel}, {Max}, {Bauman}, {Olivier}, {Wizinowich}, \& {Acton}}]{Patience2002}
{Patience}, J., {White}, R.~J., {Ghez}, A.~M., {et~al.} 2002, \apj, 581, 654

\bibitem[{{Petit} {et~al.}(2014){Petit}, {Louge}, {Th{\'e}ado}, {Paletou},
  {Manset}, {Morin}, {Marsden}, \& {Jeffers}}]{Petit2014}
{Petit}, P., {Louge}, T., {Th{\'e}ado}, S., {et~al.} 2014, \pasp, 126, 469

\bibitem[{{Rodler} {et~al.}(2013{\natexlab{a}}){Rodler}, {K{\"u}rster}, \&
  {Barnes}}]{Rodler2013}
{Rodler}, F., {K{\"u}rster}, M., \& {Barnes}, J.~R. 2013{\natexlab{a}}, \mnras,
  432, 1980

\bibitem[{{Rodler} {et~al.}(2008){Rodler}, {K{\"u}rster}, \&
  {Henning}}]{Rodler2008}
{Rodler}, F., {K{\"u}rster}, M., \& {Henning}, T. 2008, \aap, 485, 859

\bibitem[{{Rodler} {et~al.}(2010){Rodler}, {K{\"u}rster}, \&
  {Henning}}]{Rodler2010}
{Rodler}, F., {K{\"u}rster}, M., \& {Henning}, T. 2010, \aap, 514, A23

\bibitem[{{Rodler} {et~al.}(2013{\natexlab{b}}){Rodler}, {K{\"u}rster},
  {L{\'o}pez-Morales}, \& {Ribas}}]{Rodler2013b}
{Rodler}, F., {K{\"u}rster}, M., {L{\'o}pez-Morales}, M., \& {Ribas}, I.
  2013{\natexlab{b}}, Astronomische Nachrichten, 334, 188

\bibitem[{{Rodler} {et~al.}(2012){Rodler}, {Lopez-Morales}, \&
  {Ribas}}]{Rodler2012}
{Rodler}, F., {Lopez-Morales}, M., \& {Ribas}, I. 2012, \apjl, 753, L25

\bibitem[{{Rowe} {et~al.}(2008){Rowe}, {Matthews}, {Seager}, {Miller-Ricci},
  {Sasselov}, {Kuschnig}, {Guenther}, {Moffat}, {Rucinski}, {Walker}, \&
  {Weiss}}]{Rowe2008}
{Rowe}, J.~F., {Matthews}, J.~M., {Seager}, S., {et~al.} 2008, \apj, 689, 1345

\bibitem[{{Santerne} {et~al.}(2011){Santerne}, {Bonomo}, {H{\'e}brard},
  {Deleuil}, {Moutou}, {Almenara}, {Bouchy}, \& {D{\'{\i}}az}}]{Santerne2011}
{Santerne}, A., {Bonomo}, A.~S., {H{\'e}brard}, G., {et~al.} 2011, \aap, 536,
  A70

\bibitem[{{Schneider}(2011)}]{Schneider2011}
{Schneider}, J. 2011, in EPSC-DPS Joint Meeting 2011, 3

\bibitem[{{Seager}(2010)}]{Seager2010}
{Seager}, S. 2010, {Exoplanet Atmospheres: Physical Processes}

\bibitem[{{Shporer} \& {Hu}(2015)}]{Shporer2015}
{Shporer}, A. \& {Hu}, R. 2015, \aj, 150, 112

\bibitem[{{Shporer} {et~al.}(2014){Shporer}, {O'Rourke}, {Knutson},
  {Szab{\'o}}, {Zhao}, {Burrows}, {Fortney}, {Agol}, {Cowan}, {Desert},
  {Howard}, {Isaacson}, {Lewis}, {Showman}, \& {Todorov}}]{Shporer2014}
{Shporer}, A., {O'Rourke}, J.~G., {Knutson}, H.~A., {et~al.} 2014, \apj, 788,
  92

\bibitem[{{Snellen} {et~al.}(2010{\natexlab{a}}){Snellen}, {de Kok}, {de
  Mooij}, \& {Albrecht}}]{Snellen2010}
{Snellen}, I.~A.~G., {de Kok}, R.~J., {de Mooij}, E.~J.~W., \& {Albrecht}, S.
  2010{\natexlab{a}}, \nat, 465, 1049

\bibitem[{{Snellen} {et~al.}(2009){Snellen}, {de Mooij}, \&
  {Albrecht}}]{Snellen2009}
{Snellen}, I.~A.~G., {de Mooij}, E.~J.~W., \& {Albrecht}, S. 2009, \nat, 459,
  543

\bibitem[{{Snellen} {et~al.}(2010{\natexlab{b}}){Snellen}, {de Mooij}, \&
  {Burrows}}]{Snellen2010b}
{Snellen}, I.~A.~G., {de Mooij}, E.~J.~W., \& {Burrows}, A. 2010{\natexlab{b}},
  \aap, 513, A76

\bibitem[{{Sozzetti} {et~al.}(2013){Sozzetti}, {Desidera}, {Bonomo}, {Gratton},
  {Boccato}, {Claudi}, {Cosentino}, {Covino}, {Lanza}, {Maggio}, {Micela},
  {Molinari}, {Pagano}, {Piotto}, {Poretti}, \& {Smareglia}}]{Sozzetti2013}
{Sozzetti}, A., {Desidera}, S., {Bonomo}, A.~S., {et~al.} 2013, European
  Planetary Science Congress 2013, held 8-13 September in London, UK.~Online
  at: <A href=``http://meetings.copernicus.org/epsc2013''>
  http://meetings.copernicus.org/epsc2013</A>, id.EPSC2013-940, 8, EPSC2013

\bibitem[{{Sudarsky} {et~al.}(2000){Sudarsky}, {Burrows}, \&
  {Pinto}}]{Sudarsky2000}
{Sudarsky}, D., {Burrows}, A., \& {Pinto}, P. 2000, \apj, 538, 885

\bibitem[{{Takeda} {et~al.}(2007){Takeda}, {Ford}, {Sills}, {Rasio}, {Fischer},
  \& {Valenti}}]{Takeda2007}
{Takeda}, G., {Ford}, E.~B., {Sills}, A., {et~al.} 2007, \apjs, 168, 297

\bibitem[{{Valenti} \& {Fischer}(2005)}]{Valenti2005}
{Valenti}, J.~A. \& {Fischer}, D.~A. 2005, \apjs, 159, 141

\bibitem[{{van Belle} \& {von Braun}(2009)}]{Belle2009}
{van Belle}, G.~T. \& {von Braun}, K. 2009, \apj, 694, 1085

\bibitem[{{Wagener} {et~al.}(1986){Wagener}, {Caldwell}, \&
  {Fricke}}]{Wagener1986}
{Wagener}, R., {Caldwell}, J., \& {Fricke}, K.-H. 1986, \icarus, 67, 281

\bibitem[{{Walker} {et~al.}(1986){Walker}, {Diego}, {Charalambous}, {Hirst}, \&
  {Fish}}]{Walker1986}
{Walker}, D.~D., {Diego}, F., {Charalambous}, A., {Hirst}, C.~J., \& {Fish},
  A.~C. 1986, in \procspie, Vol. 627, Instrumentation in astronomy VI, ed.
  D.~L. {Crawford}, 291--302

\bibitem[{{Walker} {et~al.}(2008){Walker}, {Croll}, {Matthews}, {Kuschnig},
  {Huber}, {Weiss}, {Shkolnik}, {Rucinski}, {Guenther}, {Moffat}, \&
  {Sasselov}}]{Walker2008}
{Walker}, G.~A.~H., {Croll}, B., {Matthews}, J.~M., {et~al.} 2008, \aap, 482,
  691

\end{thebibliography}
\begin{appendix}

\section{Albedo measurements of exoplanets}

In relating the contrast to the underlying physical parameters albedo, planet radius and orbital distance, we adopt the notation and terminology used by \citet{Charbonneau1999} and \citet{Haswell2010}. The contrast $\epsilon$ is defined as the ratio of the starlight reflected by the planet to the total flux of the star as a function of wavelength and phase angle $\alpha$,
\begin{equation}
F_p(\lambda) = \epsilon(\lambda) F_*(\lambda) \Phi_{\alpha}(\lambda),
\end{equation}
where $F_*(\lambda)$ and $F_p(\lambda)$ are the observed spectra of the star and planet, and $\Phi_{\alpha}(\lambda)$ is the phase function which describes what fraction of light is scattered into any direction as determined by the global scattering properties of the surface and the atmosphere. $\alpha$ is the phase angle at which the planet is observed, and is defined as:

\begin{equation}
	\cos\left(\alpha\right) = -\sin(i)\sin\left(2\pi\varphi-\frac{1}{2}\pi\right),
\end{equation}

where $\varphi$ is the orbital phase of the planet between 0 and 1 and $i$ is the orbital inclination. $\Phi$ equals unity for a fully illuminated planet seen face-on ($\alpha = 0$) and less than unity for larger phase angles.

$\epsilon$ depends on the geometric albedo $p(\lambda)$, the radius $R_p$ and the orbital distance $a$ of the planet and is  typically on the order of $10^{-5}$ for hot Jupiters:

\begin{equation}\label{Eq:Geometric_albedo}
F_p(\lambda) = p(\lambda)\left( \frac{R_p}{a} \right)^2 \Phi_{\alpha}(\lambda) F_*(\lambda).  
\end{equation}

The geometric albedo is defined as the fraction of incident light that is scattered back into the direction of the star, compared to that of an isotropically scattering (Lambertian) disk with the same cross-section as the planet\footnote{Eq. \ref{Eq:Geometric_albedo} follows from the formal definition of the geometric albedo, which is counter-intuitive considering the fact that $p$ can be greater than $1.0$. See \citet{Seager2010} for a thorough explanation of the geometric albedo.}. The geometric albedo is not the same as the Bond albedo - which is defined as the fraction of the incident stellar energy that is scattered back into space, and which is a crucial factor for determining the equilibrium temperature of the planet. Although important, the Bond albedo is technically very challenging to determine because it requires measurements of the reflective properties of a planet at all phase angles and wavelengths. Instead, the geometric albedo and the phase function can be measured directly, and both can be used to estimate the Bond albedo. Throughout this paper, albedo refers to the wavelength-dependent geometric albedo unless explicitly stated otherwise.
\end{appendix}

\end{document}